\documentclass[aps,prr,twocolumn,superscriptaddress,longbibliography]{revtex4-2}

\usepackage[T1]{fontenc}
\usepackage[utf8]{inputenc}
\usepackage[english]{babel}
\usepackage{graphicx}
\usepackage[linkcolor=black,colorlinks=true]{hyperref}
\usepackage{amsmath,amsthm,amssymb}
\usepackage{bm}
\usepackage{physics}
\usepackage{bbold} 
\usepackage[version=4]{mhchem} 

\newcommand{\peq}{\mathrel{\phantom{=}}}
\newcommand{\ndagger}{\vphantom{\dagger}}
\newcommand{\ncc}{\vphantom{*}}
\newcommand{\bk}{\bm{k}}
\newcommand{\bmd}{\bm{d}}
\newcommand{\hbs}{\hat{\bm{\sigma}}}
\newcommand{\bnu}{\bm{\nu}}
\newcommand{\hbx}{\hat{\bm{x}}}

\newcommand{\hbz}{\hat{\bm{z}}}
\newcommand{\Eb}{\mathcal{E}_b}
\newcommand{\sign}{\mathrel{\mathrm{sign}}}

\graphicspath{{./images/}}
\allowdisplaybreaks 

\begin{document}

\title{Topological states of multiband superconductors with interband pairing}

\author{Maximilian F. Holst}
\affiliation{Institute for Theoretical Physics, ETH Zurich, 8093 Zurich, Switzerland}
\author{Manfred Sigrist}
\affiliation{Institute for Theoretical Physics, ETH Zurich, 8093 Zurich, Switzerland}
\author{Kirill V. Samokhin}
\email{kirill.samokhin@brocku.ca}
\affiliation{Department of Physics, Brock University, St. Catharines, Ontario, Canada L2S 3A1}

\begin{abstract}
We study the effects of interband pairing in two-band $s$-wave and $d$-wave superconductors with $\mathbf{D}_{4h}$ symmetry in both time-reversal invariant as well as time-reversal symmetry breaking states.
The presence of interband pairing qualitatively changes the nodal structure of the superconductor: nodes can (dis)appear, merge, and leave high-symmetry locations when interband pairing is tuned.
Furthermore, in the $d$-wave case, we find that also the boundary modes change qualitatively when interband pairing increases: flat zero-energy Andreev bound states gap out and transition to helical edge states.
\end{abstract}

\maketitle


\section{Introduction}
\label{sec: Intro}

The properties of multiband, in particular two-band, superconductors (SCs) have recently emerged as a subject of substantial interest in condensed matter physics. 
Starting with the discovery of superconductivity in \ce{MgB2} \cite{nagamatsu:2001,budko:2015}, the list of SCs in which multi-band or multi-orbital effects play an important role has been steadily growing and now includes numerous materials, such as nickel borocarbides \cite{canfield:1998}, \ce{Sr2RuO4} \cite{mackenzie:2003,agterberg:1997,kallin:2009}, \ce{NbSe2} \cite{boaknin:2003}, the heavy-fermion compounds \ce{CeCoIn5} \cite{tanatar:2005} and \ce{CePt3Si} \cite{bauer:2004}, iron-based SCs \cite{norman:2008,hirschfeld:2011}, doped topological insulators \cite{wray:2011,fu:2010}, and others.

Theoretically, a two-band generalization of the Bardeed-Cooper-Schrieffer (BCS) model was introduced in Refs.~\cite{suhl:1959,moskalenko:1959}. 
Under the assumption that the Cooper pairs are formed by the quasiparticles in the same band, i.e. intraband Cooper pairs, the order parameter in a one-dimensional (1D) pairing channel, such as $s$-wave or $d$-wave, has two components, $\eta_1$ and $\eta_2$, which describe the pairing state in each of the two bands. 
The interband scattering of the Cooper pairs between the bands couples the two order parameters as $\eta_1^*\eta_2^{\ncc}+\mathrm{c.c.}$ in the lowest order within a Ginzburg-Landau (GL) expansion, analogous to the Josephson tunneling. 
Depending on the sign of the scattering matrix element (the coefficient of the GL coupling term), the relative phase between $\eta_1$ and $\eta_2$ in the uniform ground state is either $0$ or $\pi$, corresponding to a time-reversal (TR) invariant combination. 
Subsequent studies have shown that the most significant qualitative differences from the single-band case are connected with the spatial and temporal variations of the relative phase, which produce such novel features as the Leggett modes \cite{leggett:1966}, phase solitons \cite{tanaka:2001}, and fractional vortices \cite{babaev:2002} (see for a review Ref. \cite{tanaka:2015}). 

Recent experimental and theoretical developments have motivated a further extension of the standard theory of multiband superconductivity, by taking into account the pairing among quasiparticles from different bands, i.e. interband pairing. 
Within the BCS approach of pairing in the momentum space, interband pairing is feasible if the pairing interaction cutoff energy exceeds the band splitting. 
Alternatively, starting from a real-space pairing interaction involving different atomic orbitals in a crystalline lattice, we find interband pairing components after transformation into the band representation \cite{moreo:2009,fischer:2013,ramires:2016,nomoto:2016,nica:2017}, or interband pairs arise by the proximity effect \cite{zhu:2016}. 
Assuming that interband pairs are stabilized through a suitable microscopic mechanism, one can characterize their condensate by an additional order parameter component. 
Thus, a complete phenomenological description of a two-band SC involves a GL free energy which depends on three complex order parameters, two intraband ones, $\eta_1$ and $\eta_2$, and one interband one, $\tilde\eta$. 
This increases the number of possible stable superconducting states, some of them breaking TR symmetry \cite{stanev:2010,maiti:2013,yerin:2017}.

In this paper, we show how the interband pairing affects the topological properties of a two-band SC, which is manifested in a qualitative reconstruction of the energy gap of the Bogoliubov excitations. 
We focus on two 1D pairing channels, $s$-wave and $d$-wave, on a two-dimensional (2D) square lattice and consider both TR-invariant and TR symmetry-breaking superconducting states. 
The gap functions corresponding to the intraband and interband pairing are introduced using a symmetry-based phenomenological approach. 
This approach allows one to determine the gap structure, in particular, the location of the gap nodes, even if the microscopic pairing mechanism is not known, and has proved to be very useful in the studies of unconventional fermionic superfuilds and superconductors \cite{sigrist:1991,mineev:1999}.  

According to the bulk-boundary correspondence principle, changes in the topology of the bulk state are reflected in the spectrum of the fermionic modes at the boundary \cite{volovik:2009,bernevig:2013}. 
In particular, the boundary modes are expected to be different for nodeless (fully gapped) and nodal (gapless) superconducting states. 
These boundary modes, also known as the Andreev bound states (ABSs), have been extensively used in experimental probes to identify unconventional pairing states \cite{kashiwaya:2000,sauls:2018}. 
In our study, we calculate the boundary mode spectrum by solving numerically the Bogoliubov-de Gennes (BdG) equations for a 2D lattice model of a two-band SC, and show how varying the strength of the interband pairing causes the system to undergo a series of topological phase transitions.

The paper is organized as follows:
In Sec.~\ref{sec: Gap symmetry-general}, we derive the possible interband pairing gap functions compatible with $s$-wave, $d_{xy}$-wave, and $d_{x^2-y^2}$-wave intraband pairing, respectively.
In Sec.~\ref{sec: Bulk spectrum}, we discuss the bulk spectrum and, in particular, the movement of the gap nodes in the Brillouin zone when tuning the interband pairing.
In Sec.~\ref{sec: ABS}, we numerically compute the edge spectrum of a $d_{xy}$-wave superconductor with a strip geometry and find a topological phase transition driven by the interband pairing strength.
Finally, in Sec.~\ref{sec: topology}, we analyse the topological phase found in Sec.~\ref{sec: ABS} and calculate the corresponding topological invariant(s).

Throughout the paper we use the units in which $\hbar = 1$, neglecting, in particular, the difference between the quasiparticle wave vector and momentum. Additionally, the lattice constant is set to unity.

\section{Gap symmetry: general considerations}
\label{sec: Gap symmetry-general}

We focus on a quasi-2D centrosymmetric time-reversal (TR) invariant crystal described by the point group $\mathbb{G}=\mathbf{D}_{4h}$ (however, our results can be straightforwardly generalized to other crystal symmetries); $g\in\mathbb{G}$ is either a proper rotation $R\in SO(3)$ or an improper rotation $IR$, where $I$ denotes spatial inversion.
The electron Bloch states are twofold degenerate at each wave vector $\bk = (k_x, k_y)$ due to the combined symmetry $KI$, called conjugation \cite{kittel:1987}. 
We use the index $n$ to label the bands and an additional Kramers index $s$ to distinguish two orthonormal conjugate states within the same band. 

We further assume that only two bands $n = 1, 2$ cross the chemical potential and participate in superconductivity, and also that, despite the presence of the electron-lattice spin-orbit coupling, the Bloch states in both bands transform under the point-group operations and TR in the same way as the pure spin-$1/2$ states. 
Then, the conjugacy index $s = \uparrow, \downarrow$ can be regarded as a pseudospin projection transforming under time-reversal as $K\ket{\bk, n \uparrow} = \ket{-\bk, n \downarrow}$ and $K\ket{\bk, n \downarrow} = -\ket{-\bk, n \uparrow}$, and we have
\begin{equation}
    g\ket{\bk, n s} = \sum_{s'}\ket{g\bk, n s'} D^{(1/2)}_{s's}(R).
\label{eq:bloch_basis_symmetry_action}
\end{equation}
Here, $\hat D^{(1/2)}(R)$ is the spin-$1/2$ representation of $R$. 
In other words, we assume that both bands correspond to the double-valued irreducible representation (irrep) $\Gamma_6^+$ of $\mathbf{D}_{4h}$ \cite{lax:2001}. 
The assumption \eqref{eq:bloch_basis_symmetry_action}, which is widely used in the theory of unconventional superconductivity \cite{ueda:1985}, can be relaxed and the band symmetries corresponding to other, non-pseudospin, double-valued irreps of the point group can be considered, with important consequences for the superconducting gap structure \cite{samokhin:2019, samokhin:2020}.

The superconducting system is described by the Hamiltonian
\begin{equation}
    \mathcal{H} = \mathcal{H}_0 + \mathcal{H}_{\mathrm{sc}},
\label{eq:mean_field_hamiltonian}
\end{equation}
where $\mathcal{H}_0$ is the single-particle Hamiltonian and $\mathcal{H}_{\mathrm{sc}}$ the attractive two-particle interaction Hamiltonian within a mean-field approximation.
The single-particle Hamiltonian is given by
\begin{equation*}
    \mathcal{H}_0 = \sum_{\bk n s}\xi_{\vphantom{\bk}n}^{\ndagger}(\bk)c_{\bk, n s}^{\dagger}c_{\bk, n s}^{\ndagger},
\end{equation*}
where $\xi_n(\bk) = \xi_n(-\bk)$ are the band dispersions counted from the chemical potential, so that $\xi_1(\bk) < \xi_2(\bk)$ for all $\bk$ between the two Fermi surfaces.
The superconducting mean-field pairing Hamiltonian can be represented in the following form:
\begin{equation}
    \mathcal{H}_{\mathrm{sc}} = \frac{1}{2}\sum_{\bk n n' s s'}\Delta_{nn'ss'}^{\ndagger}(\bk)c_{\bk, n s}^{\dagger}\tilde{c}_{\bk, n' s'}^{\dagger} + \mathrm{H.c.},
\label{eq:mean_field_interaction}
\end{equation}
where
\begin{equation*}
    \tilde{c}_{\bk, n s}^{\dagger} \equiv Kc_{\bk, n s}^{\dagger}K^{-1} = (i\sigma_2)_{\vphantom{\bk}s\bar{s}}^{\ndagger} c_{-\bk, n \bar{s}}^{\dagger}
\end{equation*}
are the creation operators in the time-reversed states. 

The intraband pairing in the $n$th band is described by $\hat{\Delta}_{nn}$, whereas $\hat{\Delta}_{12}$ and $\hat{\Delta}_{21}$ describe the pairing of quasiparticles from different bands (the interband pairing). 
In order to have a non-vanishing interband pairing in a BCS-like model, one has to assume that the pairing interaction shells near the Fermi surfaces, which are defined by $\abs{\xi_1},\abs{\xi_2}\leq\epsilon_c$, overlap, i.e., the pairing energy cutoff $\epsilon_c$ exceeds the typical band splitting $\Eb$. 
We do not attempt to derive the pairing Hamiltonian \eqref{eq:mean_field_interaction} from any microscopic model and regard the gap functions as phenomenological parameters.

Note that the gap functions $\hat{\Delta}_{nn'}(\bk)$ are defined in Eq.~\eqref{eq:mean_field_interaction} as the measures of the pairing between the quasiparticles in the states $\ket{\bk, n s}$ and $K\ket{\bk, n' s'}$, not between $\ket{\bk, n s}$ and $\ket{-\bk, n' s'}$. 
The gap function matrices can be represented as 
\begin{equation}
    \hat\Delta_{nn'}(\bk) = \psi_{nn'}(\bk)\hat{\sigma}_0 + \bmd_{nn'}(\bk)\hbs,
\label{eq:gap_function}
\end{equation}
where $\hat{\sigma}_0$ and $\hbs$ are respectively the unit matrix and the Pauli matrices in the pseudospin space, then Eq.~\eqref{eq:mean_field_interaction} takes the form
\begin{equation*}
\begin{split}
    \mathcal{H}_{\mathrm{sc}} 
    &= \frac{1}{2}\sum_{\bk n n' s s'}[\psi_{nn'}(\bk)(i\sigma_2)_{ss'}\\
    &\peq + \bmd_{nn'}^{\ndagger}(\bk)(i{\bm\sigma}\sigma_2)_{ss'}^{\ndagger}]c_{\bk, n s}^{\dagger}c_{-\bk, n' s'}^{\dagger} + \mathrm{H.c.}
\end{split}
\end{equation*}
Therefore, $\psi_{nn'}$ can be interpreted as the pseudospin-singlet component of the gap function and $\bmd_{nn'}$ as the pseudospin-triplet component. 
It follows from the anticommutation of the fermionic operators that $\psi_{nn'}(\bk) = \psi_{n'n}(-\bk)$ and $\bmd_{nn'}(\bk) = -\bmd_{n'n}(-\bk)$.

Additional symmetry constraints on the gap functions are obtained by looking at the transformation of the mean-field Hamiltonian \eqref{eq:mean_field_interaction} under the point-group operations and TR.
Different pairing channels correspond to different single-valued irreps $\gamma$ of $\mathbb{G}$. 
For $\mathbb{G} = \mathbf{D}_{4h}$, we consider only three even-parity pairing channels: the `$s$-wave' pairing which corresponds to the trivial irrep $A_{1g}$ and the `$d$-wave' pairing which corresponds to either $B_{1g}$ ($d_{x^2-y^2}$ pairing) or $B_{2g}$ ($d_{xy}$ pairing). 
It follows from Eq.~\eqref{eq:bloch_basis_symmetry_action} that if the pairing is described by a 1D irrep, then the gap functions satisfy the following constraints:
\begin{equation}
    \hat{D}^{(1/2)}(R)\hat{\Delta}_{nn'}(g^{-1}\bk)\hat{D}^{(1/2), \dagger}(R) = \chi_{\gamma}(g)\hat{\Delta}_{nn'}(\bk),
\label{eq:gap_function_symmetry_constraint}
\end{equation}
where $\chi_\gamma(g)$ are the group characters of the irrep $\gamma$. 
In particular, setting $g = I$, we have $\hat{\Delta}_{nn'}(-\bk) = \hat{\Delta}_{nn'}(\bk)$ because $\chi_\gamma(I)=1$ in the even irreps. 
The response of the gap functions to TR is given by $\hat{\Delta}_{nn'}^{\ndagger}(\bk)\to\hat{\Delta}_{n' n}^{\dagger}(\bk)$. 

Next, we introduce the order parameter components $\eta_{nn'}$ and represent the gap functions in the form $\hat\Delta_{nn'}(\bk) = \eta_{nn'}\hat{\phi}_{nn'}(\bk)$. 
The basis functions $\hat{\phi}_{nn'}$ which determine the momentum dependence of the gap---in particular, the location of the gap nodes---are $2\times 2$ matrices in the pseudospin space, which satisfy the point-group constraint Eq.~\eqref{eq:gap_function_symmetry_constraint} and can have singlet and triplet components similar to Eq.~\eqref{eq:gap_function}. 
Note that $\hat{\phi}_{21}(\bk) = \hat{\sigma}_2\hat{\phi}_{12}^{\top}(-\bk)\hat{\sigma}_2$ due to the anticommutation of the fermionic operators. 
Regarding the constraint imposed by TR, one can prove that the basis functions can be chosen to satisfy $\hat{\phi}_{nn'}(\bk) = \hat{\phi}_{n'n}^{\dagger}(\bk)$.

Denoting the intraband order parameters as $\eta_n\equiv\eta_{nn}$ and observing that the interband gap functions $\hat{\Delta}_{12}$ and $\hat{\Delta}_{21}$ are not independent and characterized by the same order parameter $\tilde{\eta}\equiv\eta_{12} = \eta_{21}$, we finally obtain 
\begin{equation}
\begin{split}
    \hat{\Delta}_{11}(\bk) &= \eta_1\alpha_1(\bk)\hat{\sigma}_0=\psi_1(\bk)\hat{\sigma}_0, \\
    \hat{\Delta}_{22}(\bk) &= \eta_2\alpha_2(\bk)\hat{\sigma}_0=\psi_2(\bk)\hat{\sigma}_0, \\
    \hat{\Delta}_{12}(\bk) &= \tilde{\eta}[\tilde{\alpha}(\bk)\hat{\sigma}_0 + i\tilde{\bm{\beta}}(\bk)\cdot\hbs], \\
    \hat{\Delta}_{21}(\bk) &= \tilde{\eta}[\tilde{\alpha}(\bk)\hat{\sigma}_0 - i\tilde{\bm{\beta}}(\bk)\cdot\hbs].
\end{split}
\end{equation}
Here $\alpha_{1}$, $\alpha_{2}$, $\tilde{\alpha}$ and $\tilde{\bm{\beta}}$ are real even functions of $\bk$. 
The intraband pairing in the even channels is purely singlet; the interband pairing has both singlet and triplet components. 
The Pauli principle is not violated because the exchange of electrons in an interband pair involves not only the reversal of their pseudospins and momenta but also the exchange of the band indices.

Our system is characterized by three order parameter components $\eta_1$, $\eta_2$, and $\tilde{\eta}$ which can be found by minimizing the Ginzburg-Landau free energy. 
It is easy to show that the action of TR on the order parameter components $\eta_{1}$, $\eta_{2}$ and $\tilde{\eta}$ is equivalent to complex conjugation, see Appendix \ref{app: TR-invariance}. 
One can always choose one of the components, say $\tilde{\eta}$, to be real and positive; then, $\eta_1$ and $\eta_2$ are either both real (positive or negative), which corresponds to a TR-invariant superconducting state, or have complex phases other than $0$ or $\pi$, which corresponds to a TR symmetry-breaking superconducting state, see Appendix \ref{app: OP phases}.

The point-group constraints on the basis functions take the following form
\begin{equation}
\begin{split}
    \alpha_n(g^{-1}\bk) &= \chi_{\gamma}(g)\alpha_n(\bk), \\
    \tilde{\alpha}(g^{-1}\bk) &= \chi_{\gamma}(g)\tilde{\alpha}(\bk), \\
    R(g)\tilde{\bm{\beta}}(g^{-1}\bk) &= \chi_{\gamma}(g)\tilde{\bm{\beta}}(\bk),
\label{eq:alpha_beta_symmetry_transformation}
\end{split}
\end{equation}
where $R(g)$ denotes the rotational part of $g$. 
Below we are looking for real and even-in-$\bk$ solutions of these equations, for $g=C_{4z}$ and $C_{2y}$ (the two rotational generators of the group $\mathbf{D}_{4h}$). 
To facilitate the numerical analysis later in the paper, the solutions are expressed in terms of the lattice-adapted basis functions of the even 1D irreps of $\mathbf{D}_{4h}$, namely
\begin{equation*}
\begin{split}
    f_{A_{1g}}(\bk) &= 1, \\
    f_{A_{2g}}(\bk) &= \sin(k_x)\sin(k_y)(\cos(k_x)-\cos(k_y)), \\
    f_{B_{1g}}(\bk) &= \cos(k_x) - \cos(k_y), \\
    f_{B_{2g}}(\bk) &= \sin(k_x)\sin(k_y).
\end{split}
\end{equation*}
For analytical calculations, it is more convenient to use the expressions that depend only on the direction of the wave vector in the $xy$ plane: 
\begin{equation}
\begin{split}
    f_{A_{1g}}(\bk) &= 1, \\
    f_{A_{2g}}(\bk) &= \sin(4\theta), \\
    f_{B_{1g}}(\bk) &= \cos(2\theta), \\
    f_{B_{2g}}(\bk) &= \sin(2\theta),
\end{split}
\label{eq:basis-functions-theta}
\end{equation}
where $\bk=k(\cos\theta,\sin\theta)$.

\subsection{$s$-wave pairing}
\label{sec: s-wave}

For $\gamma=A_{1g}$, the simplest singlet solutions of the symmetry constraints \eqref{eq:alpha_beta_symmetry_transformation} are given by $\alpha_1(\bk) = \alpha_2(\bk) = \tilde{\alpha}(\bk) = f_{A_{1g}} = 1$. 
Since $C_{2z}$ is a symmetry element, for the triplet interband component we have $\tilde{\bm{\beta}}(\bk) = C_{2z}\tilde{\bm{\beta}}(C_{2z}^{-1}\bk) = C_{2z}\tilde{\bm{\beta}}(-\bk) = C_{2z}\tilde{\bm{\beta}}(\bk)$; therefore, $\tilde{\beta}_1=\tilde{\beta}_2 = 0$. 
It is easy to show that $\tilde{\beta}_3\propto f_{A_{2g}}$: since $\hbz$ and $\hat{\sigma}_3$ also correspond to $A_{2g}$, $\tilde{\beta}_3\hat{\sigma}_3$ indeed corresponds to $A_{2g}\times A_{2g} = A_{1g}$. 
Collecting everything together, we arrive at the following expressions for the gap functions:
\begin{equation}
\begin{split}
    \hat{\Delta}_{11}(\bk) &= \eta_1\hat{\sigma}_0, \\ 
    \hat{\Delta}_{22}(\bk) &= \eta_2\hat{\sigma}_0, \\
    \hat{\Delta}_{12}(\bk) &= \tilde{\eta}[\hat{\sigma}_0 + i\rho f_{A_{2g}}(\bk)\hat{\sigma}_3], \\
    \hat{\Delta}_{21}(\bk) &= \tilde{\eta}[\hat{\sigma}_0 - i\rho f_{A_{2g}}(\bk)\hat{\sigma}_3],
\end{split}
\label{eq:gap_function_s_wave}
\end{equation}
where $\rho$ is a real parameter. One can say that the interband gap functions in the $A_{1g}$ channel correspond to an $s+ig$ pairing, with the understanding that the $s$ and $g$ components have a different pseudospin structure.

\subsection{$d$-wave pairing}
\label{sec: d-wave}

The singlet components of the gap functions can be chosen in the standard form: $\alpha_1(\bk) = \alpha_2(\bk) = \tilde{\alpha}(\bk) = f_{B_{1g}}(\bk)$ or $f_{B_{2g}}(\bk)$, for $d_{x^2-y^2}$- or $d_{xy}$-wave pairing, respectively. 
For the same reason as in the $s$-wave case, $\tilde{\beta}_1 = \tilde{\beta}_2 = 0$, and, using $B_{2g}\times A_{2g}=B_{1g}$ and $B_{2g}\times A_{2g}=B_{1g}$, we obtain:
\begin{equation}
\begin{split}
    \hat{\Delta}_{11}(\bk) &= \eta_1f_{B_{1g}}(\bk)\hat{\sigma}_0, \\
    \hat{\Delta}_{22}(\bk) &= \eta_2f_{B_{1g}}(\bk)\hat{\sigma}_0, \\
    \hat{\Delta}_{12}(\bk) &= \tilde{\eta}[f_{B_{1g}}(\bk)\hat{\sigma}_0 + i\rho f_{B_{2g}}(\bk)\hat{\sigma}_3], \\
    \hat{\Delta}_{21}(\bk) &= \tilde{\eta}[f_{B_{1g}}(\bk)\hat{\sigma}_0 - i\rho f_{B_{2g}}(\bk)\hat{\sigma}_3],
\end{split}
\label{eq:gap_function_dx2y2_wave}
\end{equation}
for the $d_{x^2-y^2}$-wave pairing and
\begin{equation}
\begin{split}
    \hat{\Delta}_{11}(\bk) &= \eta_1f_{B_{2g}}(\bk)\hat{\sigma}_0, \\
    \hat{\Delta}_{22}(\bk) &= \eta_2f_{B_{2g}}(\bk)\hat{\sigma}_0, \\
    \hat{\Delta}_{12}(\bk) &= \tilde{\eta}[f_{B_{2g}}(\bk)\hat{\sigma}_0 + i\rho f_{B_{1g}}(\bk)\hat{\sigma}_3], \\
    \hat{\Delta}_{21}(\bk) &= \tilde{\eta}[f_{B_{2g}}(\bk)\hat{\sigma}_0 - i\rho f_{B_{1g}}(\bk)\hat{\sigma}_3],
\end{split}
\label{eq:gap_function_dxy_wave}
\end{equation}
for the $d_{xy}$-wave pairing. 
In both cases, $\rho$ is a real parameter. 

We would like to add two comments about the structure of the interband gap functions. First, the momentum dependence of their singlet and triplet components corresponds to different even irreps of the point group (for instance, in the case of $d_{xy}$ pairing, it is $B_{2g}$ for $\tilde\alpha$ and $B_{1g}$ for $\tilde\beta_3$). 
However, the pseudospin also transforms under the point group operations, in such a way that both $\tilde\alpha$ and $\tilde\beta\hat\sigma_3$ corresponds to the same pairing channel. It is in this sense that the interband gap functions in both $B_{1g}$ and $B_{2g}$ channels correspond to a $d+id$ pairing.
Second, the singlet components in the interband gap functions in Eqs.~\eqref{eq:gap_function_dx2y2_wave} and \eqref{eq:gap_function_dxy_wave} have the same symmetry as in the intraband ones. 
As seen from Eq.~\eqref{eq:gap_function_symmetry_constraint}, this is a consequence of our assumption that both bands correspond to the same double-valued irrep of $\mathbf{D}_{4h}$. 
In general, i.e., for the bands corresponding to different irreps, the symmetry of $\tilde\alpha$ may be different from that of $\alpha$.

\section{Bogoliubov spectrum in the bulk}
\label{sec: Bulk spectrum}

The mean-field Hamiltonian Eq.~\eqref{eq:mean_field_hamiltonian} can be written in the form
\begin{equation}
    \mathcal{H} = \mathrm{const} + \frac{1}{2}\sum_{\bk}\mathcal{C}^{\dagger}(\bk)\hat{H}_{\mathrm{BdG}}(\bk)\mathcal{C}(\bk),
\label{eq:bdg_hamiltonian}
\end{equation}
where we introduced the Nambu spinor operator
\begin{equation}
    \mathcal{C}^{\top}(\bk) 
    = \left(c_{\bk, 1 \uparrow}^{\ndagger}, c_{\bk, 1 \downarrow}^{\ndagger}, \tilde{c}_{\bk, 1 \uparrow}^{\dagger}, \tilde{c}_{\bk, 1 \downarrow}^{\dagger}, c_{\bk, 2 \uparrow}^{\ndagger}, c_{\bk, 2 \downarrow}^{\ndagger}, \tilde{c}_{\bk, 2 \uparrow}^{\dagger}, \tilde{c}_{\bk, 2 \downarrow}^{\dagger}\right)
\label{eq:Nambu-operator}
\end{equation}
and the Bogoliubov--de Gennes (BdG) Hamiltonian
\begin{equation}
    \hat{H}_{\mathrm{BdG}} = \begin{pmatrix}
        \xi_1^{\ndagger}\hat{\sigma}_0^{\ndagger} & \hat{\Delta}_{11}^{\ndagger} & 0 & \hat{\Delta}_{12}^{\ndagger} \\
        \hat{\Delta}_{11}^{\dagger} & -\xi_1^{\ndagger}\hat{\sigma}_0^{\ndagger} & \hat{\Delta}_{21}^\dagger & 0 \\
        0 & \hat{\Delta}_{21}^{\ndagger} & \xi_2^{\ndagger}\hat{\sigma}_0^{\ndagger} & \hat{\Delta}_{22}^{\ndagger} \\
        \hat{\Delta}_{12}^{\dagger} & 0 & \hat{\Delta}_{22}^{\dagger} & -\xi_2^{\ndagger}\hat{\sigma}_0^{\ndagger}
    \end{pmatrix},
\label{eq:bdg_matrix}
\end{equation}
which is an $8\times 8$ matrix in the tensor product of the band, Nambu, and pseudospin spaces. 
The gap functions $\hat{\Delta}_{nn'}(\bk)$ are given by Eqs.~\eqref{eq:gap_function_s_wave}, \eqref{eq:gap_function_dx2y2_wave}, or \eqref{eq:gap_function_dxy_wave}. 

The Hamiltonian \eqref{eq:bdg_matrix} is even in $\bk$ and has the built-in particle-hole symmetry:
\begin{equation*}
    \hat{H}_{\mathrm{BdG}}(\bk) = -\hat{\cal U}_C^\dagger\hat{H}^*_{\mathrm{BdG}}(-\bk)\hat{\cal U}_C,
\end{equation*}
where
\begin{equation*}
    \hat{\cal U}_C = \begin{pmatrix}
    \hat\tau_2\otimes\hat\sigma_2 & 0 \\
    0 & \hat\tau_2\otimes\hat\sigma_2 \\
    \end{pmatrix}
\end{equation*}
and $\hat{\bm{\tau}}$ are the the Pauli matrices in the Nambu space. Since $\hat{\cal U}_C^\top=\hat{\cal U}_C$, the Hamiltonian is generically in the tenfold class D (Refs.~\cite{altland:1997,kitaev:2009,ryu:2010}). 
The TR action on the Nambu operators Eq.~\eqref{eq:Nambu-operator} is given by $K{\cal C}(\bk)K^{-1}=\hat{\cal U}_K{\cal C}(-\bk)$, where 
\begin{equation*}
    \hat{\cal U}_K = \begin{pmatrix}
    \hat\tau_0\otimes i\hat\sigma_2 & 0 \\
    0 & \hat\tau_0\otimes i\hat\sigma_2 \\
    \end{pmatrix}.
\end{equation*}
Therefore,
\begin{equation}
    K: \hat{H}_{\mathrm{BdG}}(\bk)\to\hat{\cal U}_K^\dagger\hat{H}^*_{\mathrm{BdG}}(-\bk)\hat{\cal U}_K,
\label{eq:TR-action-H-BdG}
\end{equation}
which is equivalent to replacing $(\eta_1,\eta_2,\tilde\eta)\to(\eta_1^*,\eta_2^*,\tilde\eta^*)$.
If the superconducting state is TR invariant, i.e., $\eta_1$, $\eta_2$, and $\tilde\eta$ are all real, then the BdG Hamiltonian satisfies an additional constraint:
\begin{equation*}
    \hat{H}_{\mathrm{BdG}}(\bk) = \hat{\cal U}_K^\dagger\hat{H}^*_{\mathrm{BdG}}(-\bk)\hat{\cal U}_K.
\end{equation*}
Since $\hat{\cal U}_K^\top=-\hat{\cal U}_K$, the TR invariant BdG Hamiltonian is in the tenfold class DIII. 

For the pairing symmetries we consider, the Hamiltonian Eq.~\eqref{eq:mean_field_hamiltonian} is invariant under an arbitrary $U(1)$ pseudospin rotation $c_{\bk n\uparrow(\downarrow)}^\dagger\to e^{\mp i\theta/2}c_{\bk n\uparrow(\downarrow)}^\dagger$, and we have $\comm{\hat{H}_{\mathrm{BdG}}}{\hat{\Sigma}_3} = 0$, where $\hat{\Sigma}_3 = \mathbb{1}_{4\times 4}\otimes\hat{\sigma}_3$. 
Therefore, Eq.~\eqref{eq:bdg_matrix} can be represented in the form 
$$
    \hat{H}_{\mathrm{BdG}}(\bk) = \hat{H}_{\uparrow}(\bk)\oplus\hat{H}_{\downarrow}(\bk),
$$
where the Hamiltonians
\begin{equation}
    \hat{H}_{\uparrow(\downarrow)} = \begin{pmatrix}
        \xi_1^{\ncc} & \eta_1^{\ncc}\alpha_1^{\ncc} & 0 & \tilde{\eta}(\tilde{\alpha}\pm i\tilde{\beta}) \\
        \eta_1^*\alpha_1^{\ncc} & -\xi_1^{\ncc} & \tilde{\eta}(\tilde{\alpha}\pm i\tilde{\beta}) & 0 \\
        0 & \tilde{\eta}(\tilde{\alpha}\mp i\tilde{\beta}) & \xi_2^{\ncc} & \eta_2^{\ncc}\alpha_2^{\ncc} \\
        \tilde{\eta}(\tilde{\alpha}\mp i\tilde{\beta}) & 0 & \eta_2^*\alpha_2^{\ncc} & -\xi_2^{\ncc} 
    \end{pmatrix}
\label{eq:bdg_matrix_pseudospin_split}
\end{equation}
act in the two four-dimensional eigenspaces of $\hat{\Sigma}_3$, corresponding to the two pseudospin projections. 
In Eq.~\eqref{eq:bdg_matrix_pseudospin_split} and everywhere below, we use the notation $\tilde{\beta} = \tilde{\beta}_3$ and the interband order parameter $\tilde{\eta}$ is chosen to be real and positive, but the phases of $\eta_1$ and $\eta_2$ can be arbitrary. 

It follows from Eq.~\eqref{eq:TR-action-H-BdG} that the pseudospin-resolved Hamiltonians $\hat{H}_{\uparrow}$ and $\hat{H}_{\downarrow}$ are transformed into each other by TR:
\begin{equation}
    K: \hat{H}_{\uparrow}(\bk)\to \hat{H}^*_{\downarrow}(-\bk)=\hat{H}^*_{\downarrow}(\bk).
\label{eq:H-up-h_down-TR}
\end{equation}
Also, they satisfy the relation $\hat{U}^\dagger\hat{H}_{\uparrow}^*(\bk)\hat{U}=-\hat{H}_{\downarrow}(\bk)$, where
$\hat{U} = \mathbb{1}_{2\times 2}\otimes\hat{\tau}_2$.
Introducing the magnitude and the phase of the interband gap functions
\begin{equation}
    \tilde{\eta}(\tilde{\alpha} + i\tilde{\beta}) \equiv \tilde{\Delta}(\bk) = |\tilde{\Delta}(\bk)|e^{i\tilde{\varphi}(\bk)},
\label{eq:interband_amplitude_phase_splitting}
\end{equation} 
with $|\tilde{\Delta}(\bk)|=\tilde{\eta}\tilde g(\bk)$ and $\tilde g=\sqrt{\tilde{\alpha}^2 + \tilde{\beta}^2}$, one can see that $\hat{H}_{\uparrow}$ and $\hat{H}_{\downarrow}$ are particle-hole symmetric at each $\bk$, in the following sense: 
\begin{equation*}
    \hat{U}_{C, \uparrow(\downarrow)}^{\dagger}(\bk)\hat{H}_{\uparrow(\downarrow)}^*(\bk)\hat{U}_{C, \uparrow(\downarrow)}(\bk) = -\hat{H}_{\uparrow(\downarrow)}(\bk)
\end{equation*}
where
\begin{equation*}
    \hat{U}_{C, \uparrow(\downarrow)} = \begin{pmatrix}
        e^{\mp i\tilde{\varphi}}\hat{\tau}_2 & 0 \\
	0 & e^{\pm i\tilde{\varphi}}\hat{\tau}_2
    \end{pmatrix}.
\end{equation*}
Therefore, $\hat{H}_{\uparrow}$ and $\hat{H}_{\downarrow}$ have the same bulk spectrum, which consists of symmetric pairs of eigenstates $E$ and $-E$.
Since $\hat{U}_{C, \uparrow(\downarrow)}^\top=-\hat{U}_{C, \uparrow(\downarrow)}$, the pseudospin-resolved Hamiltonians $\hat{H}_{\uparrow}$ and $\hat{H}_{\downarrow}$ are in the tenfold class C.

The matrices $\hat{H}_{\uparrow}$ and $\hat{H}_{\downarrow}$ can be diagonalized analytically, see Appendix \ref{app: calculation-E-pm}, and we find that the bulk Bogoliubov spectrum consists of four branches $\pm E_{\pm}$, where
\begin{equation}
    E_{\pm}(\bk) = \sqrt{P(\bk)\pm\sqrt{P^2(\bk)-Q^2(\bk)}} = E_{\pm}(-\bk).
\label{eq:bdg_eigenvalues}
\end{equation}
The notations are as follows:
\begin{equation*}
\begin{split}
    P &= \frac{1}{2}\left(\xi_1^2 + \abs{\psi_1}^2 + \xi_2^2 + \abs{\psi_2}^2\right) + |\tilde{\Delta}|^2, \\
    Q^2 &= r_1^2 + r_2^2 + r_3^2,
\end{split}
\end{equation*}
$\psi_n(\bk) = \eta_n\alpha_n(\bk)$ are the intraband gap functions and
\begin{equation*}
\begin{split}
    r_1 &= \xi_1\xi_2 - \abs{\psi_1\psi_2} + |\tilde{\Delta}|^2, \\
    r_2 &= \xi_1\abs{\psi_2} + \xi_2\abs{\psi_1}, \\
    r_3 &= \sqrt{2|\tilde{\Delta}|^2[\abs{\psi_1\psi_2} - \Re(\psi_1\psi_2)]}.
\end{split}
\end{equation*}
One can show that $P > Q$ in the presence of interband pairing. Therefore, $E_+$ is strictly greater than $E_-$ at all $\bk$.  
Each of the four branches $\pm E_{\pm}$ is twofold degenerate due to pseudospin. 
In the absence of interband pairing, we set $\tilde{\eta} = 0$ and recover the usual expressions for a two-band superconductor: 
\begin{equation*}
\begin{split}
    E_+(\bk) &= \max\{\epsilon_1(\bk), \epsilon_2(\bk)\}, \\
    E_-(\bk) &= \min\{\epsilon_1(\bk), \epsilon_2(\bk)\}, 
\end{split}
\end{equation*}
where $\epsilon_n = \sqrt{\xi_n^2 + \abs{\psi_n}^2}$ is the excitation energy in the $n$th band.

While the upper Bogoliubov excitation branch $E_+$ is fully gapped in the superconducting state, the lower branch $E_-$ vanishes at the wave vector $\bk$ if 
\begin{equation}
    r_1(\bk) = r_2(\bk) = r_3(\bk) = 0,
\label{eq:node_condition}
\end{equation}
in which case $E_-(\bk)$ and $-E_-(\bk)$ touch, producing a gap node. 
In two spatial dimensions, the three real functions $r_{1, 2, 3}$ cannot simultaneously vanish at the same $\bk$, unless forced to do so by additional symmetries. 

Writing the intraband order parameters in the form 
\begin{equation}
    \eta_1 = \abs{\eta_1}e^{i\varphi_1},\quad \eta_2 = \abs{\eta_2}e^{i\varphi_2},
\label{eq:eta_1-eta_2-modulus-phase}
\end{equation}
we see that $r_3$ identically vanishes in the states in which $\varphi_1 + \varphi_2 = 0$ or $2\pi$. 
This happens, in particular, in the TR invariant states in which $\eta_1$ and $\eta_2$ are both either real positive or real negative. 
As shown in Appendix \ref{app: OP phases}, the TR-symmetry breaking states with $\varphi_1 + \varphi_2 = 0$ or $2\pi$ are stable only if the system's parameters are fine-tuned, the possibility that can be neglected. 
In a generic state with the interband pairing, i.e., when $\tilde{\eta}\neq 0$ and $\varphi_1 + \varphi_2\neq 0$ or $2\pi$, $r_3(\bk)=0$ only if $\psi_1(\bk) = 0$ or $\psi_2(\bk) = 0$. 
In the $s$-wave case, this can only happen accidentally and is neglected. 
In contrast, the $d$-wave intraband gap functions, and therefore $r_3$, vanish along the high-symmetry directions for symmetry reasons. Thus, there exist four classes of the stable bulk nodal structures, which are studied below. 

\subsection{Generic $s$-wave pairing}
\label{sec: s-wave-Phi-nonzero}

In this case, the phases of $\eta_1$ and $\eta_2$ take any values, except $\varphi_1=\varphi_2=0$ or $\pi$. The TR invariant states in which $\eta_1$ and $\eta_2$ are real but have opposite signs are also included here. Since $r_3$ is nonzero at all $\bk$, the $s$-wave superconducting state is fully gapped, regardless of the strength of the interband pairing.

\subsection{Generic $d$-wave pairing}
\label{sec: d-wave-Phi-nonzero}

\begin{figure}[!t]
    \centering
    \includegraphics[width=\linewidth]{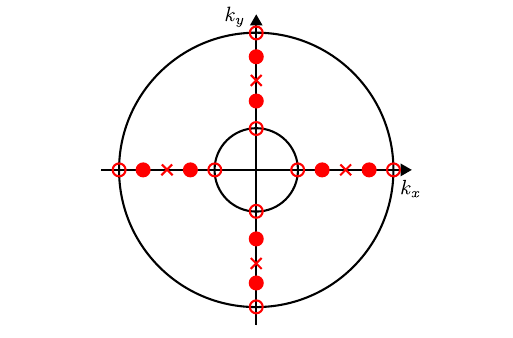}
    \caption{%
    Schematic illustration of the nodal behavior in the toy model of the generic $d_{xy}$-wave case (Sec.~\ref{sec: d-wave-Phi-nonzero}).
    Empty red dots: high-symmetry nodes without interband pairing $\tilde{\eta} = 0$.
    Filled red dots: high-symmetry nodes at small interband pairing $0 < \tilde{\eta} < \tilde{\eta}_{c}$, see Eq.~\eqref{eq:eta-c-generic-d-wave}.
    Red crosses: annihilation of high-symmetry nodes at interband pairing $\tilde{\eta} = \tilde{\eta}_{c}$.
    }
    \label{fig:toy_model_nodes_1_generic_dwave}
\end{figure}

For concreteness, let us consider the evolution of the nodal structure in the $d_{xy}$-wave case (for the $d_{x^2-y^2}$-wave pairing, the nodes are just rotated by $\pi/4$). 
In the absence of the interband pairing, $r_3$ identically vanishes and the point gap nodes are located where $\xi_1 = \psi_1 = 0$ or $\xi_2 = \psi_2 = 0$, i.e., at the intersections of the axes of the 2D Brillouin zone with the Fermi surfaces.

In the presence of the interband pairing and for generic phases of $\eta_1$ and $\eta_2$, $r_3$ only vanishes along the axes $k_x = 0$ and $k_y = 0$ for symmetry reasons. 
Moreover, $r_2$ is also zero there and the only remaining gap node condition, Eq.~\eqref{eq:node_condition}, takes the following form:
\begin{equation}
    \xi_1\xi_2 = -\tilde\eta^2\tilde g^2
\label{eq:node_equations_d_wave}
\end{equation}
along the $k_x = 0$ or $k_y = 0$ lines. 
As $\tilde{\eta}$ increases, the nodes remain on the high-symmetry axes, but move into the ``interband space'', where $\xi_1 < 0$ and $\xi_2 > 0$  (recall that we assume $\xi_1 < \xi_2$). 
Eventually, at a sufficiently strong interband pairing, the nodes merge and annihilate each other, which marks the transition into a fully gapped bulk phase, as shown in Fig.~\ref{fig:toy_model_nodes_1_generic_dwave}. 
Annihilating nodes were also found in Ref.~\cite{chubukov:2016}, in a model of a TR-invariant SC with a $d_{x^2-y^2}$-wave intraband pairing and $d_{xy}$-wave interband pairing.

The evolution of the gap structure can be studied analytically using a simple model with two parabolic electron-like bands:
\begin{equation}
    \xi_{1(2)}(\bk)=\xi(\bk)\mp\frac{\Eb}{2},\quad \xi(\bk)=\frac{k^2-k_0^2}{2m},
\label{eq:simple-model-bands}
\end{equation}
where $\Eb>0$ is the band splitting, in which the two Fermi surfaces are circles of radii $k_{F,1(2)}=k_0\sqrt{1\pm m\Eb/k_0^2}$. We use the following gap symmetry factors:
\begin{equation}
\begin{split}
    & \alpha_1 = \alpha_2=\tilde\alpha=\sin(2\theta),\\ & \tilde\beta = \rho\cos(2\theta),
\end{split}
\label{eq:simple-model-d-wave-functions}
\end{equation}
see Eqs.~\eqref{eq:gap_function_dxy_wave} and \eqref{eq:basis-functions-theta}. 
The gap node equation Eq.~\eqref{eq:node_equations_d_wave} becomes
\begin{equation*}
    \xi^2=\left(\frac{\Eb}{2}\right)^2-\tilde\eta^2\rho^2,
\end{equation*}
along the axes of the momentum space. Taking, for instance, the $\theta=0$ axis, at $\tilde\eta=0$ the two nodes are located on the Fermi surfaces, at $\bk=k_{F,1}\hbx$ and $\bk=k_{F,2}\hbx$. As $\tilde\eta$ increases, the nodes move towards each other, to $\bk=k_1\hbx$ and $\bk=k_2\hbx$, where
\begin{equation}
    k_{1,2}=k_0\left[1\pm\frac{2m}{k_0^2}\sqrt{\left(\frac{\Eb}{2}\right)^2-\tilde\eta^2\rho^2}\right]^{1/2}.
\label{eq:nodes-k1-k2}
\end{equation}
Finally, when the interband order parameter $\tilde\eta$ reaches the critical value
\begin{equation}
    \tilde\eta_c=\frac{\Eb}{2|\rho|},
\label{eq:eta-c-generic-d-wave}
\end{equation}
the nodes merge at $\bk=k_0\hbx$ and ``annihilate'' each other. At stronger interband pairing, our $d_{xy}$-wave superconductor is fully gapped. 
Note that the disappearance of the nodes happens only if $\rho\neq 0$, i.e., when the interband gap functions contain the triplet component with the $d_{x^2-y^2}$-wave-like momentum dependence.

\subsection{$s$-wave pairing, $\varphi_1 = \varphi_2 = 0$ or $\pi$}
\label{sec: s-wave-Phi-zero}

\begin{figure}[!t]
    \centering
    \includegraphics[width=\linewidth]{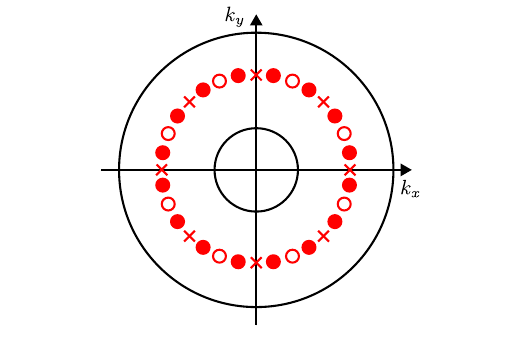}
    \caption{%
    Schematic illustration of the nodal behavior in the toy model of the exceptional $s$-wave case (Sec.~\ref{sec: s-wave-Phi-zero}).
    Empty red dots: stray nodes at interband pairing $\tilde{\eta} = \tilde{\eta}_{c, 1}$, see Eq.~\eqref{eq:eta-c1-s-wave}.
    Filled red dots: stray nodes at interband pairing $\tilde{\eta}_{c, 1} < \tilde{\eta} < \tilde{\eta}_{c, 2}$, see Eq.~\eqref{eq:eta-c2-s-wave}.
    Red crosses: annihilation of stray nodes at interband pairing $\tilde{\eta} = \tilde{\eta}_{c, 2}$.
    }
    \label{fig:toy_model_nodes_2_exceptional_swave}
\end{figure}

In this TR invariant state, $r_3$ is identically zero everywhere, but we still need to solve the remaining equations $r_1=0$ and $r_2 = 0$. Assuming, without loss of generality, the same intraband symmetry factors in both bands: $\alpha_1(\bk)=\alpha_2(\bk)=\alpha(\bk)$, the equation $r_2 = 0$ takes the form $|\alpha|(\xi_1|\eta_2|+\xi_2|\eta_1|)=0$.
One way to satisfy this is to put $\alpha=0$, but that does not happen in the $s$-wave case, whereas in the $d$-wave case that can happen only along the high-symmetry lines, which was already considered in Sec.~\ref{sec: d-wave-Phi-nonzero}. Therefore, if we look for the gap nodes away from the high-symmetry axes, then we need to solve the following two equations: 
\begin{equation}
\begin{split}
    \abs{\xi_1(\bk)\xi_2(\bk)} &= \tilde\eta^2\tilde g^2(\bk)-\abs{\eta_1\eta_2}\alpha^2(\bk), \\
    0 &= \xi_1(\bk)\abs{\eta_2}+\xi_2(\bk)\abs{\eta_1}.
\end{split}
\label{eq:node_equations_exceptional-TRI}
\end{equation}
Note that the second equation can have solutions only between the two Fermi surfaces, where $\xi_1 < 0$ and $\xi_2 > 0$, i.e., $\xi_1\xi_2=-\abs{\xi_1\xi_2}$.
If the interband pairing is sufficiently strong, so that the right-hand side of the first equation is positive, then Eq.~\eqref{eq:node_equations_exceptional-TRI} defines two lines between the Fermi surfaces. The intersections of these lines, if they exist, correspond to accidental point nodes in the excitation spectrum.

To illustrate these points for the $s$-wave pairing, we use the band structure model Eq.~\eqref{eq:simple-model-bands}, with the following angular dependence of the gap functions:
\begin{equation*}
    \alpha=\tilde\alpha=1,\quad \tilde\beta=\rho\sin(4\theta),
\end{equation*}
see Eqs.~\eqref{eq:gap_function_s_wave} and \eqref{eq:basis-functions-theta}. 
Solving Eq.~\eqref{eq:node_equations_exceptional-TRI}, we obtain that the accidental nodes are located on the circle of the radius 
\begin{equation}
    K=k_0\sqrt{1+\zeta\frac{m\Eb}{k_0^2}},\quad k_{F,2}<K<k_{F,1},
\label{eq:nodal-circle-s-wave}
\end{equation}
where
\begin{equation*}
    \zeta=\frac{\abs{\eta_2}-\abs{\eta_1}}{\abs{\eta_2}+\abs{\eta_1}},\quad \abs{\zeta}<1,
\end{equation*}
at the angles found from the equation
\begin{equation*}
    \tilde\eta^2[1+\rho^2\sin^2(4\theta)]-|\eta_1\eta_2|=(1-\zeta^2)\left(\frac{\Eb}{2}\right)^2.
\end{equation*}
In the absence of the interband pairing, this last equation does not have any solutions. 
As $\tilde\eta$ increases and reaches
\begin{equation}
    \tilde\eta_{c,1}=\sqrt{\frac{|\eta_1\eta_2|}{1+\rho^2}}\sqrt{1+\left(\frac{\Eb}{|\eta_1|+|\eta_2|}\right)^2},
\label{eq:eta-c1-s-wave}    
\end{equation}
the nodes emerge in pairs, first at the angles given by $\theta=\pi/8,3\pi/8,\dots$. 
As the interband pairing further increases, the nodes split and move along the circle defined by Eq.~\eqref{eq:nodal-circle-s-wave} towards the angles $\theta=0,\pi/4,\dots$, where they finally merge and disappear at 
\begin{equation}
    \tilde\eta_{c,2}=\sqrt{1+\rho^2}\,\tilde\eta_{c,1}.
\label{eq:eta-c2-s-wave}    
\end{equation}
These changes in the gap structure are shown in Fig.~\ref{fig:toy_model_nodes_2_exceptional_swave}. 
The superconducting state is fully gapped at $\tilde\eta<\tilde\eta_{c,1}$ and at $\tilde\eta>\tilde\eta_{c,2}$. 

Note that the gap nodes appear in this model only if $\rho\neq 0$, i.e., when the interband gap functions contain the ``triplet'' component with an anisotropic, $g$-wave-like momentum dependence. 
Although these nodes are topologically unstable, since any deviation from the condition $\varphi_1 = \varphi_2 = 0$ or $\pi$ will remove them, they are protected by TR symmetry.  

\subsection{$d$-wave pairing, $\varphi_1 = \varphi_2 = 0$ or $\pi$}
\label{sec: d-wave-Phi-zero}

\begin{figure}[!t]
    \centering
    \includegraphics[width=\linewidth]{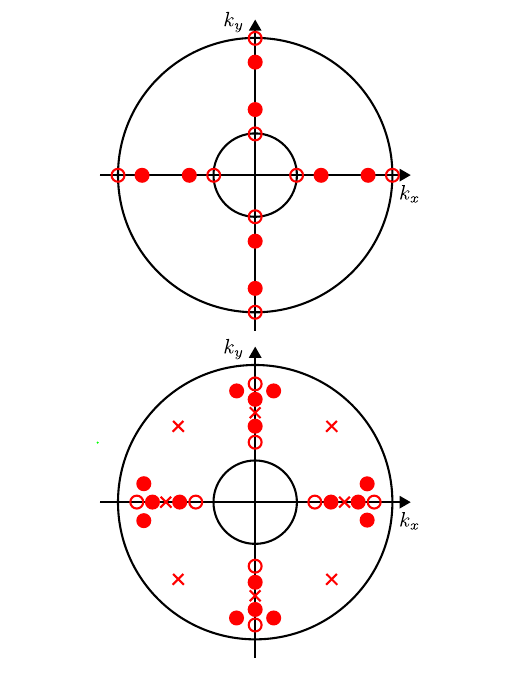}
    \caption{%
    Schematic illustration of the nodal behavior in the toy model of the exceptional $d_{xy}$-wave case (Sec.~\ref{sec: d-wave-Phi-zero}).
    (Top)
    Empty red dots: high-symmetry nodes without interband pairing $\tilde{\eta} = 0$.
    Filled red dots: high-symmetry nodes at small interband pairing $0 < \tilde{\eta} < \tilde{\eta}_{c, 1}$, see Eq.~\eqref{eq:eta-c1-d-wave-rho-1}.
    (Bottom)
    Empty red dots: high-symmetry nodes at interband pairing $\tilde{\eta}\lesssim\tilde{\eta}_{c, 1}$.
    Filled red dots: high-symmetry nodes (along the main axes) and stray nodes (off the main axes) at interband pairing $\tilde{\eta}_{c, 1} < \tilde{\eta} < \tilde{\eta}_{c, 2}$, see Eq.~\eqref{eq:eta-c2-d-wave-rho-1}.
    Red crosses: annihilation of high-symmetry nodes (along the main axes) at interband pairing $\tilde{\eta} = \tilde{\eta}_{c}$ and annihilation of stray nodes (off the main axes) at interband pairing $\tilde{\eta} = \tilde{\eta}_{c, 2}$ ($\tilde{\eta}_{c} < \tilde{\eta}_{c, 2}$).
    }
    \label{fig:toy_model_nodes_3_exceptional_dwave}
\end{figure}

The difference from the generic $d$-wave case is that $r_3$ now vanishes everywhere, which makes it possible for additional gap nodes to appear away from the high-symmetry lines. Repeating the reasoning from Sec.~\ref{sec: s-wave-Phi-zero}, we find that there are two types of nodes: the ``high-symmetry'' ones, which are located where $\alpha(\bk)=0$, and also the ``stray'' ones, which correspond to the solutions of Eq.~\eqref{eq:node_equations_exceptional-TRI}. 

To develop some analytical insight, we again assume a $d_{xy}$-wave pairing and use the parabolic bands \eqref{eq:simple-model-bands}, with the symmetry factors given by Eq.~\eqref{eq:simple-model-d-wave-functions}. 
We obtain that the stray nodes are located on the circle defined by Eq.~\eqref{eq:nodal-circle-s-wave}, at the angles determined by the equation
\begin{equation}
\begin{split}
    (1-\zeta^2)\left(\frac{\Eb}{2}\right)^2 
    &= \tilde\eta^2[\sin^2(2\theta)+\rho^2\cos^2(2\theta)] \\
    &\peq -|\eta_1\eta_2|\sin^2(2\theta).
\end{split}
\label{eq:nodal-angles-exceptional-d-wave}
\end{equation} 
At $\tilde\eta=0$, this equation has no solutions. To illustrate the different scenarios of how the stray nodes are created and destroyed by increasing the interband pairing strength, we solve Eq.~\eqref{eq:nodal-angles-exceptional-d-wave} in three cases, for $\rho=0$, $|\rho|\gg 1$, and $\rho=1$.

At $\rho=0$, which corresponds to the absence of the triplet $d_{x^2-y^2}$ component in the interband gap functions, the stray nodes appear in pairs at $\theta=\pi/4,3\pi/4,\dots$ when $\tilde\eta$ reaches the critical value
\begin{equation*}
\tilde\eta_{c,1}=\sqrt{|\eta_1\eta_2|}\sqrt{1+\left(\frac{\Eb}{|\eta_1|+|\eta_2|}\right)^2}.   
\end{equation*}
As the interband pairing strength further increases, the nodes split and move away from each other, staying on the circle (\ref{eq:nodal-circle-s-wave}) and asymptotically approaching the axes $\theta=0,\pi/2,\dots$. 
Since at $\rho=0$ both the intraband and interband gap functions vanish along the axes, the high-symmetry nodes are not affected by $\tilde\eta$, i.e., always remain at the intersections of the two Fermi surfaces with the lines $k_x=0$ and $k_y=0$. 

At $|\rho|\gg 1$, which corresponds to the triplet $d_{x^2-y^2}$ component dominating the interband gap functions, the stray nodes appear at $\theta=0,\pi/2,\dots$ when the interband pairing strength reaches
\begin{equation*}
    \tilde\eta_{c,1}=\frac{\sqrt{|\eta_1\eta_2|}}{|\rho|(|\eta_1|+|\eta_2|)}\Eb.
\end{equation*}
At this point they ``peel off'' in pairs from the high-symmetry nodes and, as $\tilde\eta$ increases, move along the circle \eqref{eq:nodal-circle-s-wave} towards the diagonals. 
Whereas the high-symmetry nodes annihilate each other at $\tilde\eta_{c}>\tilde\eta_{c,1}$, see Eq.~\eqref{eq:eta-c-generic-d-wave}, the stray nodes survive in the limit $\tilde\eta\gg\tilde\eta_{c,1}$, asymptotically approaching the axes $\theta=\pi/4,3\pi/4,\dots$.

To see what happens in the general case, when both the singlet ($d_{xy}$) and the triplet ($d_{x^2-y^2}$) components are present in the interband gap functions, we set $\rho=1$. 
Then, the solutions of Eq.~\eqref{eq:nodal-angles-exceptional-d-wave} exist only if $\tilde\eta_{c,1}\leq\tilde\eta\leq\tilde\eta_{c,2}$,
where 
\begin{equation}
    \tilde\eta_{c,1}=\frac{\sqrt{|\eta_1\eta_2|}}{|\eta_1|+|\eta_2|}\Eb,
\label{eq:eta-c1-d-wave-rho-1}    
\end{equation}
and
\begin{equation}
    \tilde\eta_{c,2}=\sqrt{|\eta_1\eta_2|}\sqrt{1+\left(\frac{\Eb}{|\eta_1|+|\eta_2|}\right)^2},
\label{eq:eta-c2-d-wave-rho-1}    
\end{equation}
As $\tilde\eta$ increases, the stray nodes first appear on the axes, i.e., at $\theta=0,\pi/2,\dots$, where they peel off in pairs from the high-symmetry nodes, then move towards $\theta=\pi/4,3\pi/4,\dots$, where they eventually merge and annihilate each other. 
Note that $\tilde\eta_{c,1}$ is less than the critical strength of the interband pairing at which the high-symmetry nodes disappear, see Eq.~\eqref{eq:eta-c-generic-d-wave}. 
Therefore, there is an interval of $\tilde\eta$, in which the stray nodes co-exist with the high-symmetry ones, so that there are sixteen nodes altogether (eight of each type), all located between the Fermi surfaces, as shown in Fig.~\ref{fig:toy_model_nodes_3_exceptional_dwave}. 
Similar behaviour of the nodes was also found in a different model in Ref.~\cite{nakayama:2018}, in which the interband pairing in a TR invariant $d$-wave state is controlled by the inter-orbital SO coupling.
In contrast to the limiting cases of $\rho=0$ and $\rho\gg 1$, in which $\tilde\eta_{c,2}=\infty$ and the stray nodes survive the strong interband pairing, in the general case all nodes eventually disappear as $\tilde\eta$ increases.

The stray nodes are accidental, in the sense that they are not protected by the crystal symmetry. 
In order to destroy them, one has to tune the intraband order parameter phases out of the TR invariance condition $\varphi_1 = \varphi_2 = 0$ or $\pi$. However, this condition always corresponds to a critical point of the free energy, see Appendix \ref{app: OP phases}, and if this critical point is a minimum, then the state is stable. Therefore, the stray nodes are protected by the TR symmetry.

\subsection{Summary}
\label{sec: nodes-summary}

The effect of the interband pairing on the energy gap nodes in the bulk is fundamentally different in the four cases discussed in this section.
The least interesting case is the generic $s$-wave state, which is fully gapped at $\tilde\eta=0$ and remains so as $\tilde\eta$ increases. 
This state is topologically trivial and does not support zero-energy boundary modes.

In the TR-invariant $s$-wave state with $\varphi_1 = \varphi_2 = 0$ or $\pi$, a sufficiently strong interband pairing can create and then destroy again point nodes between the Fermi surfaces. 
The critical values of $\tilde\eta$ separating the gapped and gapless phases, as well as the locations of the nodes, are model-dependent. 

In the generic $d$-wave state, the nodes are located only along the high-symmetry lines. 
As $\tilde\eta$ increases, these nodes leave the Fermi surfaces, move towards each other into the interband space, and merge and disappear at $\tilde\eta=\tilde\eta_c$. 
The gapless and gapped phases separated by $\tilde\eta_c$ are expected to be topologically different, which is confirmed in Secs.~\ref{sec: ABS} and \ref{sec: topology} below.

The TR-invariant $d$-wave state with $\varphi_1 = \varphi_2 = 0$ or $\pi$ exhibits the most complex behaviour. 
In this case, the nodes along the high-symmetry lines can co-exist with the additional (stray) nodes in the interband space, whose number and locations are model-dependent. 
As the interband pairing increases, the two families of nodes evolve as the system passes through a series of transitions characterized by the creation and destruction of the pairs of nodes. Across these transitions, the topological charges of the nodes are conserved, see Sec.~\ref{sec: topology-gapless}.
Eventually, at a sufficiently large $\tilde\eta$, all the nodes will have pairwise collided and annihilated each other, so that the superconducting state will be fully gapped.

\section{Boundary modes}
\label{sec: ABS}

\begin{figure*}[!t]
    \centering
    \includegraphics[width=\linewidth]{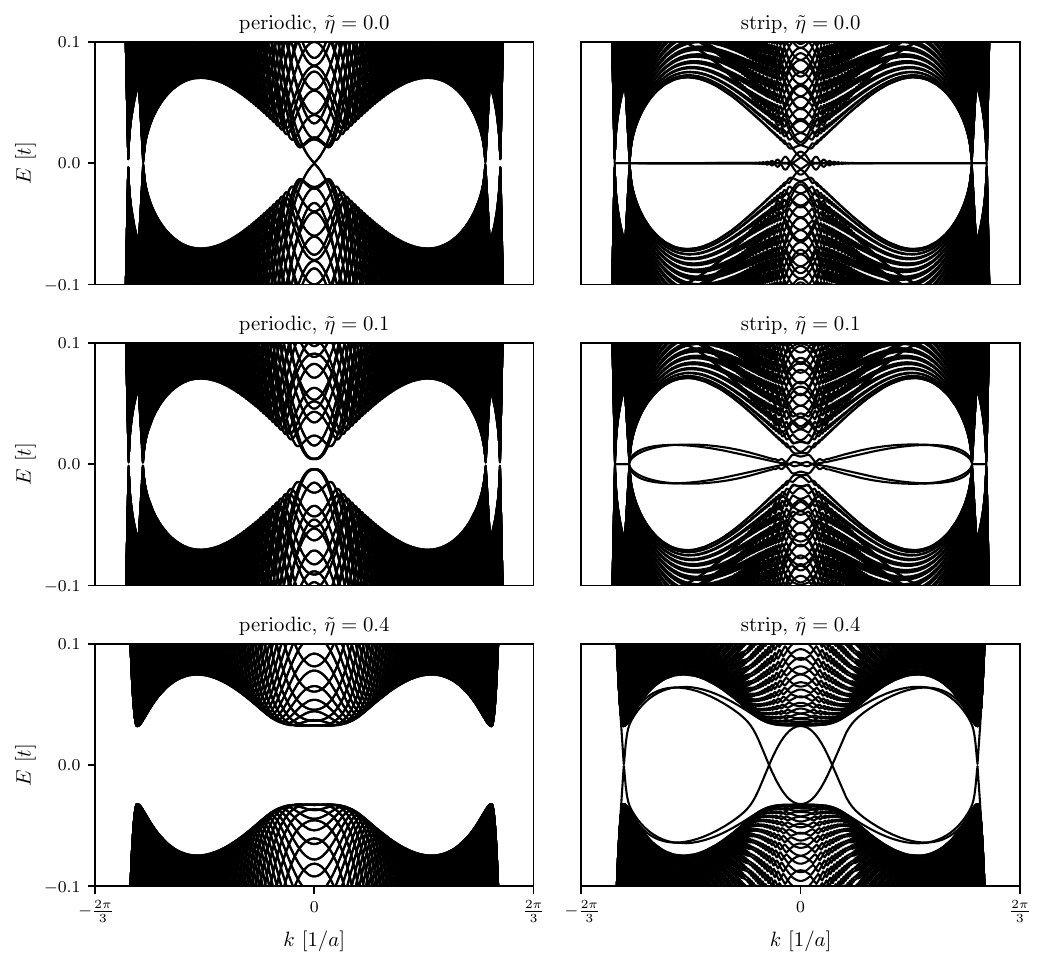}
    \caption{%
    Energy spectrum of $\hat{H}_{\uparrow}(k)$, see Eq.~\eqref{eq:strip_bdg_hamiltonian_spin_split}, in the generic $d_{xy}$-wave case, for increasing interband pairing $\tilde{\eta}$ (top to bottom). 
    Left column: periodic boundary conditions. 
    Right column: strip geometry. 
    Parameters: $N_x = 2501$, $N_y = 500$, $\mu = -1.5$, $t_1 = 1.2$, $t_2 = 0.8$, $t_1' = 0.5$, $t_2' = 0.0$, $\eta_1 = 0.11$, $\eta_2 = -0.09$, $\rho = 0.5$. 
    Energy is measured in the units of $t = (t_1+t_2)/2$.
    }
    \label{fig:periodic_vs_strip_spectrum}
\end{figure*}

\begin{figure}[!t]
    \centering
    \includegraphics[width=\columnwidth]{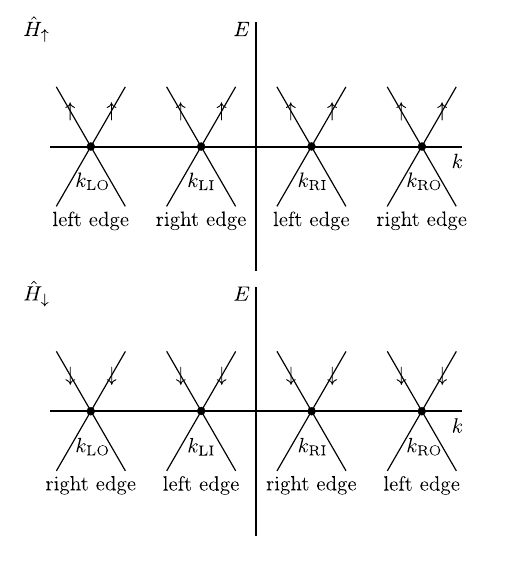}
    \caption{Schematic illustration of the edge states in a fully gapped $d_{xy}$-wave SC, for strong interband pairing (top panel: the edge states for $\hat{H}_{\uparrow}$, bottom panel: the edge states for $\hat{H}_{\downarrow}$).}
    \label{fig:edge_states}
\end{figure}

Having discussed the bulk properties of the different superconducting phases in Sec.~\ref{sec: Bulk spectrum}, we turn our attention towards the boundaries of the material.
We consider a strip geometry: the system is infinitely extending along the $x$ direction, but has a finite width along the $y$ direction.
To formulate the lattice model, we assume $N_x$ ($N_y$) lattice sites with periodic (open) boundary conditions along the $x$ ($y$) direction.
As a result, the system effectively possesses two infinitely extended edges parallel to the $x$ axis.

The bulk band structure is described by
\begin{equation*}
\begin{split}
    \xi_n(\bm{k}) &= - 2t_n\left(\cos(k_x) + \cos(k_y)\right) \\
    &\peq - 4t_n'\cos(k_x)\cos(k_y) - \mu,
\end{split}
\end{equation*}
where $\mu$ is the chemical potential, $t_n$ is the nearest, and $t_n'$ -- the next-nearest neighbor hopping amplitude in the $n$th band.
Superconductivity is either of $s$- or $d_{xy}$-wave type, given by Eq.~\eqref{eq:gap_function_s_wave} and Eq.~\eqref{eq:gap_function_dxy_wave}, respectively; for $d_{x^2-y^2}$-wave superconductivity, we would consider edges rotated by $\pi/4$.
The total bulk Hamiltonian of the lattice model is described by Eq.~\eqref{eq:mean_field_hamiltonian}.

The formal description of the superconducting strip system follows Ref.~\cite{holst:2022}.
The momentum component $k_y$ is not a good quantum number, because translation symmetry is broken along the $y$ direction.
To account for this, we only consider the momentum representation $k \equiv k_x$ along the $x$ direction, while we keep the real space representation $i \equiv i_y$ along the $y$ direction.
Assuming a sufficiently wide strip, the superconducting gap is approximately constant along the $y$ direction and we neglect any potential surface effects causing the order parameter to be spatially deformed close to the edges.
The order parameters $\eta_1$, $\eta_2$, and $\tilde{\eta}$ are not computed self-consistently, but set to their respective bulk values.

Similar to Eq.~\eqref{eq:bdg_hamiltonian}, the total mean-field Hamiltonian is of the form
\begin{equation}
    \mathcal{H} = \mathrm{const} + \frac{1}{2}\sum_{k}\mathcal{C}^{\dagger}(k)\hat{H}_{\mathrm{BdG}}(k)\mathcal{C}^{\ndagger}(k),
\label{eq:bdg_hamiltonian_2}
\end{equation}
where $\mathcal{C}^{\dagger} = \mathcal{C}_{\uparrow}^{\dagger}\oplus\mathcal{C}_{\downarrow}^{\dagger}$ and $\hat{H}_{\mathrm{BdG}} = \hat{H}_{\uparrow}\oplus\hat{H}_{\downarrow}$. The spin-resolved Nambu operators are given by
\begin{align*}
    \mathcal{C}_{s}^{\dagger}(k) &= \left(c_{k 1, 1 s}^{\dagger}, c_{k 1, 2 s}^{\dagger}, \dots, c_{k N, 2 s}^{\dagger} \right. \\
    &\peq\left.\tilde{c}_{k 1, 1 s}^{\ndagger}, \tilde{c}_{k 1, 2 s}^{\ndagger}, \dots, \tilde{c}_{k N, 2 s}^{\ndagger}\right),
\end{align*}
where $c_{k i, n s}^{\dagger}$ ($c_{k i, n s}^{\ndagger}$) creates (annihilates) an electron with momentum $k = k_x$ at position $i = i_y = 1, \dots, N_y$ in band $n = 1, 2$ with spin $s = \uparrow, \downarrow$, and
\begin{equation}
    \hat{H}_{\uparrow(\downarrow)} = \begin{pmatrix}
        \hat{\xi}_{\uparrow(\downarrow)} & \hat{\Delta}_{\uparrow(\downarrow)} \\
        \hat{\Delta}^{\dagger}_{\uparrow(\downarrow)} & -\hat{\xi}_{\uparrow(\downarrow)}
    \end{pmatrix}.
\label{eq:strip_bdg_hamiltonian_spin_split}
\end{equation}
The exact forms for the $2N_y\times 2N_y$ matrices $\hat{\xi}_{\uparrow(\downarrow)}$ and $\hat{\Delta}_{\uparrow(\downarrow)}$ are provided in Appendix \ref{app: numerical-BdG}, for both the $s$- and $d_{xy}$-wave cases. 

The BdG Hamiltonian $\hat{H}_{\mathrm{BdG}}$ is diagonal in the pseudospin space, with the blocks $\hat{H}_{\uparrow}$ and $\hat{H}_{\downarrow}$ being related by TR, see Eq.~\eqref{eq:H-up-h_down-TR}.
Furthermore, it is diagonal in $k$ space.
As a consequence, the problem is reduced to the diagonalization of a $4N_y\times 4N_y$ matrix.
Here, we employ an exact diagonalization procedure to solve for the eigenvalues as well as the corresponding eigenstates of $\hat{H}_{\uparrow}(k)$.

We begin the discussion of our results with the $s$-wave case.
For a weak interband pairing, the spectrum is fully gapped in both the generic (cf. Sec.~\ref{sec: s-wave-Phi-nonzero}) and the exceptional cases (cf. Sec.~\ref{sec: s-wave-Phi-zero}).
In the generic case, this situation remains true regardless of the strength of the interband pairing.
In contrast, in the exceptional case, there exist two critical values, $\tilde{\eta}_{c, 1}$ and $\tilde{\eta}_{c, 2}$, between which the spectrum is gapless for four distinct $k$ values (three $k$ values at $\tilde{\eta} = \tilde{\eta}_{c, 1}$ and two $k$ values at $\tilde{\eta} = \tilde{\eta}_{c, 2}$).
As the interband pairing strength increases from $\tilde{\eta}_{c, 1}$ to $\tilde{\eta}_{c, 2}$, these nodes move towards each other until they annihilate eventually, so that the spectrum is fully gapped again.
There are no edge states present---regardless of the interband pairing strength.

Next, we discuss the generic $d_{xy}$-wave case (cf. Sec.~\ref{sec: d-wave-Phi-nonzero}).
The results are summarized in Fig.~\ref{fig:periodic_vs_strip_spectrum}.
In the absence of interband pairing (the top panels), the energy spectrum shows five zeros at $k_{\mathrm{L O}} < k_{\mathrm{L I}} < k_0 \equiv 0 < k_{\mathrm{R I}} < k_{\mathrm{R O}}$ (L = `Left', R = `Right', O = `Outer', I = `Inner').
These correspond to the bulk nodes along the main axes and are located exactly on the two Fermi surfaces.
The node at $k_0$ is four-fold degenerate corresponding to the four nodes of the system along the $y$ direction.
Between $k_{\mathrm{L O}}$ and $k_{\mathrm{R O}}$ the spectrum shows flat ABSs.
As soon as the interband pairing is turned on, the ABSs between $k_{\mathrm{L I}}$ and $k_{\mathrm{R I}}$ gap out, while they remain intact between $k_{\mathrm{L O}}$ and $k_{\mathrm{L I}}$ as well as between $k_{\mathrm{R I}}$ and $k_{\mathrm{R O}}$.

As the interband pairing strength increases, the nodes move away from the Fermi surfaces until they meet each other and annihilate.
After this point, the bulk is completely gapped but eight zero-energy crossing ABS branches remain.
They are singly degenerate (doubly degenerate for $\hat{H}_{\uparrow}\oplus\hat{H}_{\downarrow}$ if both edges are taken into account, see Fig.~\ref{fig:edge_states}) and the corresponding eigenstates are localized near the edges of the strip, see Appendix \ref{app: numerical-BdG}.
They mark a different, topologically non-trivial, superconducting phase.

The edge states are schematically illustrated in Fig.~\ref{fig:edge_states}.
For $\hat{H}_{\uparrow}$, four states are located close to the left edge of the strip, while the other four are located close to the right edge of the strip.
Depending on their slope they move either along the positive or negative $x$ direction.
Furthermore, they mix electrons from one band with holes from the other band, see Appendix \ref{app: numerical-BdG}.

Finally, we turn our attention towards the exceptional $d_{xy}$-wave case (cf. Sec.~\ref{sec: d-wave-Phi-zero}).
Similarly to the generic case, we observe a topological phase transition when the interband pairing strength increases.
However, apart from the phase transition, the behavior of the spectrum significantly differs from the generic case.
Weak interband pairing does not immediately and fully gap out the ABSs.
Indeed, there are ABSs present until the stray nodes annihilate along the diagonals of the Brillouin zone (cf. Sec.~\ref{sec: d-wave-Phi-zero}).

We make two final remarks.
First, the topological phase transition occurs regardless of the presence of TR symmetry.
The characteristics of the edge states in the strip spectrum (cf. Fig.~\ref{fig:periodic_vs_strip_spectrum}) are the same regardless of the intraband order parameter phases $\varphi_{n}$.
Only the behavior of the bulk nodes and the ABSs in the gapless regime differs between the TR symmetry-breaking generic states and the TR invariant exceptional states.
Second, the interband pairing strength required to reach the topologically non-trivial superconducting phase strongly depends on the distance between the two Fermi surfaces.
The closer the Fermi surfaces along the main axes of the Brillouin zone, the weaker the required interband pairing in order to annihilate the gap nodes.

\section{Topological arguments}
\label{sec: topology}

The results of the previous two sections show that the effects of the interband pairing are most profound in the $d$-wave states. The evolution of the bulk gap structure, which is reflected in the changes of the ABS spectrum, can be interpreted in terms of a series of transitions between topologically distinct superconducting phases. In this section, we discuss the relevant bulk topological invariants, focusing as before on the $d_{xy}$-wave states.

\subsection{Gapless bulk}
\label{sec: topology-gapless}

\begin{figure}[!t]
    \centering
    \includegraphics[width=\linewidth]{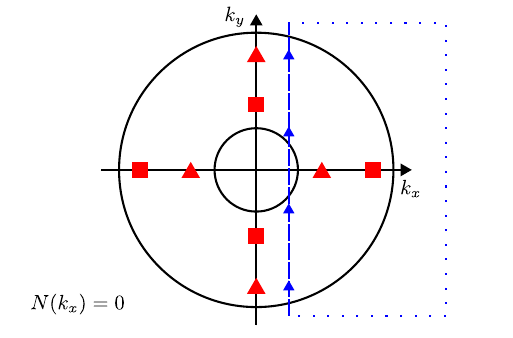}
    \caption{%
    Integration contour for computing the topological invariant $N(k_x)$, see Eq.~\eqref{eq:ZEABS-number-dxy}.
    Red triangles and squares: gap node positions at finite interband pairing $0 < \tilde{\eta} < \tilde{\eta}_c$ (generic $d_{xy}$-wave case).
    The different shapes indicate that the nodes have opposite topological charges.
    }
    \label{fig:topological_charges_1}
\end{figure}

\begin{figure}[!t]
    \centering
    \includegraphics[width=\linewidth]{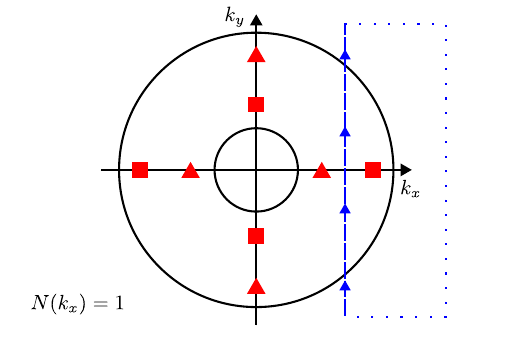}
    \caption{%
    Integration contour for computing the topological invariant $N(k_x)$, see Eq.~\eqref{eq:ZEABS-number-dxy}.
    Red triangles and squares: gap node positions at finite interband pairing $0 < \tilde{\eta} < \tilde{\eta}_c$ (generic $d_{xy}$-wave case).
    The different shapes indicate that the nodes have opposite topological charges.
    }
    \label{fig:topological_charges_2}
\end{figure}

\begin{figure}[!t]
    \centering
    \includegraphics[width=\linewidth]{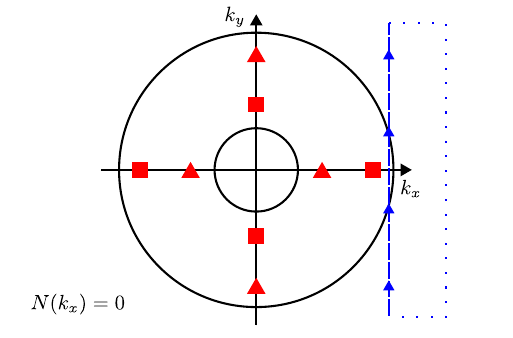}
    \caption{%
    Integration contour for computing the topological invariant $N(k_x)$, see Eq.~\eqref{eq:ZEABS-number-dxy}.
    Red triangles and squares: gap node positions at finite interband pairing $0 < \tilde{\eta} < \tilde{\eta}_c$ (generic $d_{xy}$-wave case).
    The different shapes indicate that the nodes have opposite topological charges.
    }
    \label{fig:topological_charges_3}
\end{figure}

According to Sec.~\ref{sec: Bulk spectrum}, the gap structure of a generic nodal $d_{xy}$-wave state is insensitive to the phases of the intraband order parameters: the nodes move on the high-symmetry axes as the interband pairing strength varies. 
In order to study the topological properties, one can focus on the TR invariant states, in which the order parameter components $\eta_1$, $\eta_2$, and $\tilde{\eta}$ are all real. If $\eta_1\eta_2<0$, then there exist only the high-symmetry nodes (Sec.~\ref{sec: d-wave-Phi-nonzero}), whereas at $\eta_1\eta_2>0$ the stray nodes are also possible (Sec.~\ref{sec: d-wave-Phi-zero}). 

For the real order parameters, the Hamiltonians $\hat{H}_{\uparrow}$ and $\hat{H}_{\downarrow}$, see Eq.~\eqref{eq:bdg_matrix_pseudospin_split}, have a `chiral' symmetry:
\begin{equation*}
    \hat{U}_S^{\dagger}\hat{H}_{\uparrow(\downarrow)}^{\ndagger}(\bk)\hat{U}_S^{\ndagger} = -\hat{H}_{\uparrow(\downarrow)}^{\ndagger}(\bk),\quad 
    \hat{U}_S = \begin{pmatrix}
        \hat{\tau}_2 & 0 \\
	0 & \hat{\tau}_2
    \end{pmatrix}.
\end{equation*}
In the basis in which $\hat{U}_S$ is diagonal, the Hamiltonians can be brought to a block off-diagonal form, e.g.,
\begin{equation*}
    \hat H_\uparrow(\bk)\to \hat V\hat H_\uparrow(\bk)\hat V^\dagger=
    \begin{pmatrix}
    0 & \hat\upsilon(\bk) \\
    \hat\upsilon^\dagger(\bk) & 0
    \end{pmatrix},
\end{equation*}
where
\begin{equation*}
    \hat V=\frac{1}{\sqrt{2}}
    \begin{pmatrix}
    1 & -i & 0 & 0 \\
    0 & 0 & 1 & -i \\
    -i & 1 & 0 & 0 \\
    0 & 0 & -i & 1 
    \end{pmatrix}
\end{equation*}
and
\begin{equation}
    \hat\upsilon=\begin{pmatrix}
    \psi_1+i\xi_1 & |\tilde\Delta| e^{i\tilde\varphi} \\
    |\tilde\Delta| e^{-i\tilde\varphi} & \psi_2+i\xi_2
    \end{pmatrix}.
\label{eq:chiral-block}
\end{equation}
The $\upsilon$-matrix for $\hat H_\downarrow$ is obtained from Eq.~\eqref{eq:chiral-block} by replacing $\tilde\varphi\to-\tilde\varphi$. 

The positions of the gap nodes are determined by the zeros of $\abs{\det\hat\upsilon}$, whereas the topological charges of the nodes are given by the winding number of the phase of $\det\hat\upsilon$: 
\begin{equation}
    q=\oint\frac{d\bk}{2\pi i}\bm{\nabla}_{\bk}\ln\det\hat\upsilon,
\label{eq:top-charge-winding}
\end{equation}
see Refs.~\cite{sato:2011,schnyder:2011}. 
The integration here is performed around an infinitesimally small circular contour wrapping counterclockwise around the node. 
From Eq.~\eqref{eq:chiral-block}, we have
\begin{equation}
    \det\hat\upsilon=\psi_1\psi_2-\xi_1\xi_2-|\tilde\Delta|^2+i(\xi_1\psi_2+\xi_2\psi_1),
\label{eq:det-chiral-block}
\end{equation}
where $\psi_n(\bk)=\eta_n\alpha(\bk)$, assuming the same intraband symmetry factors in both bands. 
In agreement with the results of Sec.~\ref{sec: Bulk spectrum},  we see that the zeros of $|\det\hat\upsilon|$ are located in the interband space, either where $\alpha=0$ and $\xi_1\xi_2+\tilde\Delta^2=0$ (the high-symmetry nodes), or away from the symmetry axes (the stray nodes), the latter being possible only if $\eta_1$ and $\eta_2$ have the same sign.

The topological charges of the nodes can be easily calculated by expanding $\det\hat\upsilon(\bk)$ in the vicinity of the nodes. 
In the case $\eta_1>0$, $\eta_2<0$, we find (see Appendix \ref{app: nodal charges}) that the gap nodes located on the same axis are oppositely ``charged'', as shown in Figs.~\ref{fig:topological_charges_1}, \ref{fig:topological_charges_2}, and \ref{fig:topological_charges_3}, which makes it possible for the nodes to ``annihilate'' each other, as discussed in Sec.~\ref{sec: d-wave-Phi-nonzero}. 

According to Refs.~\cite{sato:2011,schnyder:2011}, the number of the zero-energy edge modes at given momentum $k_x$ along the boundary, per one pseudospin projection, is equal to $|N(k_x)|$, where
\begin{equation*}
    N(k_x)=\Im\int_{-\infty}^\infty\frac{dk_y}{2\pi}\;\nabla_{k_y}\ln\det\hat\upsilon(\bk).
\end{equation*}
The integral here is taken along a straight line which runs between the opposite edges of the Brillouin zone perpendicular to the boundary. In a continuum model, the limits are extended to infinity. Assuming that all gap functions vanish far from the Fermi surfaces, one can integrate along a closed contour $C$ shown in Figs.~\ref{fig:topological_charges_1}, \ref{fig:topological_charges_2}, and \ref{fig:topological_charges_3}.
Using Stokes’ theorem to contract the contour without crossing any gap nodes, we find that
$N(k_x)$ is equal to the total charge of the nodes enclosed by $C$.

In this way, we obtain:
\begin{equation}
	|N(k_x)|=\left\{
	\begin{array}{ll}
	0,\quad & \mathrm{at}\ |k_x|<k_2,\\ 
	1,\quad & \mathrm{at}\ k_2<|k_x|<k_1,\\
        0,\quad & \mathrm{at}\ k_1<|k_x|,
	\end{array}
	\right.
\label{eq:ZEABS-number-dxy}	
\end{equation}
where $k_{1,2}$ are the positions of the bulk nodes, see Eq.~\eqref{eq:nodes-k1-k2}.
Taking into account the pseudospin degeneracy, the total number of the zero-energy ABS localized near one edge of the sample is equal to $2|N(k_x)|$. 
We see that the momentum range in which the topologically protected zero-energy boundary modes exist shrinks with increasing the interband pairing and eventually disappears, in agreement with the numerical results of Sec.~\ref{sec: ABS}. 

\subsection{Gapped bulk}
\label{sec: topology-gapped}

\begin{figure}[!t]
    \centering
    \includegraphics[width=\columnwidth]{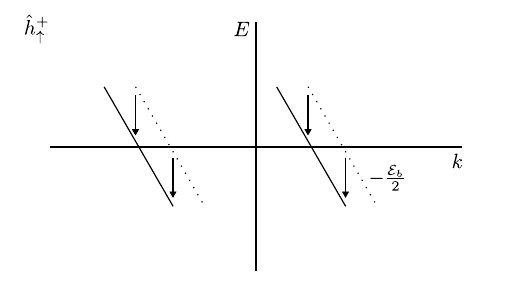}
    \caption{Schematic illustration of the chiral edge modes for the Hamiltonian $\hat{h}_{\uparrow}^{+}$, see Eq.~\eqref{eq:h-plus-minus}.
    Dashed lines: chiral modes without the energy shift.
    Solid lines: chiral modes shifted by $-\Eb/2$.}
    \label{fig:edge_states2}
\end{figure}

We have seen in the previous sections that the bulk Bogoliubov spectrum becomes fully gapped when the interband pairing exceeds certain value. Moreover, if $\tilde\Delta\neq 0$, then the Bogoliubov branches $E_+$ and $E_-$, see Eq. (\ref{eq:bdg_eigenvalues}), are always separated. In the absence of any level crossings, the intraband pairing can be adiabatically turned off without affecting the bulk topology. Therefore, in order to study the topology of the mappings $\bk\to\hat H_\uparrow(\bk)$ and $\bk\to\hat H_\downarrow(\bk)$ in the nodeless regime, we can set $\eta_1=\eta_2=0$, which considerably simplifies the calculations. The pseudospin-resolved Hamiltonians \eqref{eq:bdg_matrix_pseudospin_split} are then reduced to direct sums of $2\times 2$ matrices: 
\begin{equation*}
    \hat{H}_{\uparrow}(\bk) = \hat{h}_\uparrow^+(\bk)\oplus\hat{h}_\uparrow^-(\bk),\quad 
    \hat{H}_{\downarrow}(\bk) = \hat{h}_\downarrow^+(\bk)\oplus\hat{h}_\downarrow^-(\bk),
\end{equation*}   
where
\begin{equation}
\begin{split}
    \hat{h}_{\uparrow}^\pm &= \pm\frac{\xi_1 - \xi_2}{2}\hat{\sigma}_0 + \bnu_\pm\hbs,\\
    \hat{h}_{\downarrow}^\pm &= \mp\frac{\xi_1 - \xi_2}{2}\hat{\sigma}_0 + \bnu_\pm\hbs,
\end{split}   
\label{eq:h-plus-minus}
\end{equation}
and $\bnu_\pm = (\tilde\eta\tilde\alpha, \mp\tilde\eta\tilde\beta, \xi)$ and $\xi = (\xi_1 + \xi_2)/2$. Since the Hamiltonians $\hat{H}_{\uparrow}$ and $\hat{H}_{\downarrow}$ are in the class C, the gapped bulk states are characterized by an even ($2\mathbb{Z}$) topological invariant.\cite{ryu:2010} Therefore, we expect an even number of zero-energy boundary modes.

In the case of the $d_{xy}$-wave pairing, we have    
\begin{equation}
    \tilde\alpha\pm i\tilde\beta = \sin(2\theta)\pm i\rho\cos(2\theta),
\label{eq:d-pm-id}
\end{equation}
see Eqs.~\eqref{eq:gap_function_dxy_wave} and \eqref{eq:basis-functions-theta}. 
Therefore, the matrices \eqref{eq:h-plus-minus} have the same form as the BdG Hamiltonians for the chiral $d\pm id$ states, shifted up or down in energy. It is well known \cite{volovik:1997} that the $d+id$ and $d-id$ superconductors can support chiral boundary modes, which are protected by the bulk topology.   

Diagonalizing Eq.~\eqref{eq:h-plus-minus} and using the band model \eqref{eq:simple-model-bands}, we find that the bulk spectra of $\hat{H}_{\uparrow}$ and $\hat{H}_{\downarrow}$ are the same and, in agreement with Eq.~\eqref{eq:bdg_eigenvalues}, are given by four particle-hole symmetric branches $\pm E_\pm$, where
\begin{equation*}
    E_\pm=\left|\sqrt{\xi^2 + |\tilde\Delta|^2}\pm\frac{\Eb}{2}\right|
\end{equation*} 
and $\tilde\Delta$ has the form \eqref{eq:interband_amplitude_phase_splitting}. 
Note that the branch indices in this last expression have nothing to do with the ``chirality'' index $\pm$ in Eq.~\eqref{eq:h-plus-minus}.
At a sufficiently strong interband pairing, $|\tilde\Delta|>\Eb/2$, the bulk spectrum is fully gapped. 

According to Ref.~\cite{volovik:2009}, the topological invariant characterizing a gapped chiral $d$-wave state has the following form:
\begin{equation}
    N = \frac{1}{4\pi}\int d^2\bk\; \hat{\bnu}\left(\frac{\partial\hat{\bnu}}{\partial k_x}\times \frac{\partial\hat{\bnu}}{\partial k_y}\right),
\label{eq:N_k_integral_1}
\end{equation} 
where $\hat{\bnu} = \bnu/\abs{\bnu}$ and $\bnu=\bnu_+$ or $\bnu_-$. 
Here we integrate over the 2D momentum space, which can be compactified into an $S^2$ sphere, because the gap functions vanish outside the overlapping BCS pairing shells, see Sec.~\ref{sec: Gap symmetry-general}, so that $\hat{\bnu} = \hat{\bm z}\sign\xi$ and the integrand in Eq.~\eqref{eq:N_k_integral_1} is equal to zero far from the Fermi surfaces. The expression~\eqref{eq:N_k_integral_1} is nothing but the degree of the mapping $\bk\to\hat{\bnu}(\bk)$, which takes integer values and can be used to enumerate different equivalence classes of the Hamiltonians Eq.~\eqref{eq:h-plus-minus}.   

Writing the interband gap functions in the form Eq.~\eqref{eq:interband_amplitude_phase_splitting}, with $|\tilde\Delta|$ nonvanishing only inside the pairing shells of thickness $\epsilon_c$, Eq.~\eqref{eq:N_k_integral_1} takes the form
\begin{equation*}
    N = \frac{1}{4\pi}\int d^2\bk\, \frac{|\tilde\Delta|^2}{(\xi^2 + |\tilde\Delta|^2)^{3/2}}\left(\frac{\partial\xi}{\partial k_x}\frac{\partial\tilde\varphi}{\partial k_y} - \frac{\partial\xi}{\partial k_y}\frac{\partial\tilde\varphi}{\partial k_x}\right).
\end{equation*} 
Finally, neglecting the $\xi$-dependence of $|\tilde\Delta|$ and $\tilde\varphi$ inside the pairing shell, sending $\epsilon_c\to\infty$, and integrating with respect to $\xi$, we obtain:
\begin{equation*}
    N = \frac{1}{2\pi}\oint d\tilde\varphi,
\end{equation*} 
where the integration is performed along the $\xi = 0$ line. Since we neglect the $\xi$-dependence of $\tilde\varphi$, one could integrate along either of the two Fermi surfaces, with the same result (note that the lines $\xi=0$, $\xi_1 = 0$, and $\xi_2 = 0$ all lie within the BCS pairing shell, which encompasses both Fermi surfaces).  
Thus, the invariant \eqref{eq:N_k_integral_1} is equal to the phase winding number of the interband gap function. 

For the $d_{xy}$-wave pairing, see Eq.~\eqref{eq:d-pm-id}, we obtain the following winding numbers for $\hat{h}_{\uparrow}^\pm$ and $\hat{h}_{\downarrow}^\pm$:
\begin{equation*}
    N_\pm = \mp 2\sign\rho.
\end{equation*}
Therefore, each of the four Hamiltonians $\hat{h}_{\uparrow}^\pm$ and $\hat{h}_{\downarrow}^\pm$ has two chiral zero modes near each edge of the sample. These modes have opposite slopes for opposite chiralities, and are also shifted up and down in energy by $\pm\Eb/2$, as shown in Fig.~\ref{fig:edge_states2} for $\hat{h}_{\uparrow}^+$ and $\rho>0$. 
The $4\times 4$ Hamiltonian $\hat H_s$ corresponding to one pseudospin channel has four helical modes composed of two pairs of the counter-propagating chiral modes from $\hat{h}_s^+$ and $\hat{h}_s^-$, as shown in Fig.~\ref{fig:edge_states}. This result is in agreement with the numerical solution of the BdG equations, see Fig. \ref{fig:periodic_vs_strip_spectrum}.

\section{Conclusion}
\label{sec: Conclusion}

Based on a symmetry analysis, we determined the possible interband pairing gap functions in the case of two-band superconductors with $s$-wave, $d_{xy}$-wave, or $d_{x^2-y^2}$-wave pairing, which can give rise to both TR invariant and TR symmetry-breaking superconducting states.
As the interband pairing strength increases, the nodal structure changes fundamentally.
Nodes leave the Fermi surfaces and eventually annihilate each other on the high-symmetry axes, whereas other nodes (stray nodes) appear, move, and merge in the interband space.

In the case of a $d$-wave superconductor with a strip geometry, the boundary modes exhibit qualitative changes when interband pairing increases. 
Starting from zero-energy flat ABSs in the absence of interband pairing, these modes partially gap out as soon as interband pairing is turned on.
In the limit of strong interband pairing, the system undergoes a topological phase transition to a fully gapped helical $d\pm id$-wave superconducting state.
The corresponding topological invariant is the phase winding number of the interband gap, which explains the existence of the eight gap-crossing zero-energy branches near one edge of the sample in the helical state.

\begin{acknowledgments}
We thank M. Fischer and A. Ramires for helpful discussions. 
This work was supported by the Swiss National Science Foundation (SNSF) through Division II (No. 184739) (MH and MS) and by Discovery Grant 2021-03705 from the Natural Sciences and Engineering Research Council of Canada (KS). 
KS is grateful to the Institute for Theoretical Physics, ETH Zurich for hospitality and the Pauli Center for Theoretical Studies for financial support.
\end{acknowledgments}

\bibliography{references}

\clearpage
\appendix

\section{Response to TR}
\label{app: TR-invariance}

Suppose that the triplet component of the interband gap function is given by $\tilde{\bm{\beta}} = (0, 0, \tilde{\beta})$, see Secs.~\ref{sec: s-wave} and \ref{sec: d-wave}. 
Then, the pairing Hamiltonian Eq.~\eqref{eq:mean_field_interaction} takes the following form:
\begin{eqnarray*}
\begin{split}
    \hat{H}_{\mathrm{sc}} 
    &= \frac{1}{2}\sum_{\bk}\left[\eta_1^{\ndagger}\alpha_1^{\ndagger}(c_{\bk, 1 \uparrow}^{\dagger}\tilde{c}_{\bk, 1 \uparrow}^{\dagger} + c_{\bk, 1 \downarrow}^{\dagger}\tilde{c}_{\bk, 1 \downarrow}^\dagger)\right. \\ 
    &\peq +\eta_2^{\ndagger}\alpha_2^{\ndagger}(c_{\bk, 2 \uparrow}^{\dagger}\tilde{c}_{\bk, 2 \uparrow}^{\dagger} + c_{\bk, 2 \downarrow}^{\dagger}\tilde{c}_{\bk, 2 \downarrow}^{\dagger}) \\
    &\peq +\tilde{\eta}(\tilde{\alpha} + i\tilde{\beta})(c_{\bk, 1 \uparrow}^{\dagger}\tilde{c}_{\bk, 2 \uparrow}^{\dagger} + c_{\bk, 2 \downarrow}^{\dagger}\tilde{c}_{\bk, 1 \downarrow}^\dagger) \\
    &\peq\left. +\tilde{\eta}(\tilde{\alpha} - i\tilde{\beta})(c_{\bk, 1 \downarrow}^{\dagger}\tilde{c}_{\bk, 2 \downarrow}^{\dagger} + c_{\bk, 2 \uparrow}^\dagger\tilde{c}_{\bk, 1 \uparrow}^{\dagger})\right] + \mathrm{H.c.}.
\end{split}
\end{eqnarray*}
Since $K(c_{\bk, n s}^{\dagger}\tilde{c}_{\bk, n' s'}^{\dagger})K^{-1} = c_{\bk, n' s'}^{\dagger}\tilde{c}_{\bk, n s}^{\dagger}$, the TR-transformed Hamiltonian is given by
\begin{eqnarray*}
\begin{split}
    K\hat{H}_{\mathrm{sc}}K^{-1} 
    &= \frac{1}{2}\sum_{\bk}\left[\eta_1^{*\ndagger}\alpha_1^{\ndagger}(c_{\bk, 1 \uparrow}^{\dagger}\tilde{c}_{\bk, 1 \uparrow}^{\dagger} + c_{\bk, 1 \downarrow}^{\dagger}\tilde{c}_{\bk, 1 \downarrow}^\dagger)\right. \\ 
    &\peq +\eta_2^{*\ndagger}\alpha_2^{\ndagger}(c_{\bk, 2 \uparrow}^{\dagger}\tilde{c}_{\bk, 2 \uparrow}^{\dagger} + c_{\bk, 2 \downarrow}^{\dagger}\tilde{c}_{\bk, 2 \downarrow}^{\dagger}) \\
    &\peq +\tilde{\eta}(\tilde{\alpha} - i\tilde{\beta})(c_{\bk, 2 \uparrow}^{\dagger}\tilde{c}_{\bk, 1 \uparrow}^{\dagger} + c_{\bk, 1 \downarrow}^{\dagger}\tilde{c}_{\bk, 2 \downarrow}^\dagger) \\
    &\peq\left. +\tilde{\eta}(\tilde{\alpha} + i\tilde{\beta})(c_{\bk, 2 \downarrow}^{\dagger}\tilde{c}_{\bk, 1 \downarrow}^{\dagger} + c_{\bk, 1 \uparrow}^\dagger\tilde{c}_{\bk, 2 \uparrow}^{\dagger})\right] + \mathrm{H.c.}
\end{split}
\end{eqnarray*}
which is the same as $\hat{H}_{\mathrm{sc}}$, with $(\eta_1, \eta_2, \tilde{\eta})$ replaced by $(\eta_1^*, \eta_2^*, \tilde{\eta}^*)$. 
If the order parameter is real, then $K\hat H_{sc}K^{-1}=\hat H_{sc}$, i.e., the Hamiltonian is intrinsically complex, but TR-invariant.

\section{Stable states}
\label{app: OP phases}

The three order parameter components can be combined into $\bm{\eta} = (\eta_1, \eta_2, \tilde{\eta})^{\top}$. The second- and fourth-order uniform terms in the GL free energy density have the following form: 
\begin{equation}
    F_2 = \bm{\eta}^{\dagger}\hat{A}\bm{\eta},\quad
    \hat{A} = \left(\begin{array}{ccc}
	A_{11} & A_{12} & \tilde A_{13} \\
	A_{12} & A_{22} & \tilde A_{23} \\
	\tilde A_{13} & \tilde A_{23} & \tilde A_{33}
	\end{array}\right),
 \label{F-GL-2}
\end{equation}
where $\hat A$ is a real symmetric matrix and
\begin{equation*}
\begin{split}
    F_4 &= \beta_1\abs{\eta_1}^4 + \beta_2\abs{\eta_2}^4 + \tilde\beta_1\abs{\eta_1}^2\abs{\tilde{\eta}}^2 + \tilde{\beta}_2\abs{\eta_2}^2\abs{\tilde{\eta}}^2 \\
    &\peq + \tilde{\beta}_3\abs{\tilde{\eta}}^4 + \tilde{\beta}_4(\eta_1\eta_2\tilde{\eta}^{*, 2} + \mathrm{c.c.}),
\end{split}
\end{equation*}
see Ref.~\cite{samokhin:NA} for the microscopic derivation. 
The diagonal elements of the matrix $\hat{A}$ depend on temperature, so that $\hat{A}$ loses positive-definiteness at the critical temperature $T_c$. 
In the absence of the interband pairing, all the quantities with tildes are zero and Eq.~\eqref{F-GL-2} takes the usual form for a two-band superconductor, with $A_{12}$ describing the Josephson tunneling of the Cooper pairs between the bands.

Just below $T_c$, the order parameter is small and the quartic terms in the free energy can be neglected. Choosing $\tilde{\eta}$ to be real positive and writing the intraband order parameters in the form \eqref{eq:eta_1-eta_2-modulus-phase}, the phase-dependent terms in the energy can be represented as
\begin{equation}
    F(\varphi_1, \varphi_2) = a\cos(\varphi_1 - \varphi_2) + \tilde a_1\cos\varphi_1 + \tilde a_2\cos\varphi_2,
\label{eq:gl_phase_form}
\end{equation} 
where $a$, $\tilde{a}_1$, and $\tilde{a}_2$ are proportional to the off-diagonal elements of $\hat{A}$ and can be positive or negative. Minimizing Eq.~\eqref{eq:gl_phase_form}, we obtain:
\begin{equation}
\begin{split}
    a\sin(\varphi_1 - \varphi_2) + \tilde{a}_1\sin\varphi_1 &= 0, \\
    a\sin(\varphi_1 - \varphi_2) - \tilde{a}_2\sin\varphi_2 &= 0.
\end{split}
\label{eq:gl_phase_eqs}
\end{equation}
These equations always have four solutions $\varphi_1, \varphi_2 = 0$ or $\pi$, which correspond to the TR invariant superconducting states. 
Whether these states are stable or not depends on the parameters in Eq.~\eqref{eq:gl_phase_form}.

In general, Eq.~\eqref{eq:gl_phase_eqs} can also have solutions different from $0$ or $\pi$, which describe TR symmetry-breaking superconducting states. To construct these solutions, we employ the following procedure. First, we pick some values of $\varphi_1$ and $\varphi_2$ and use Eq.~\eqref{eq:gl_phase_eqs} to obtain:
\begin{equation}
    \tilde a_1=-a\frac{\sin(\varphi_1-\varphi_2)}{\sin\varphi_1},\quad
    \tilde a_2=a\frac{\sin(\varphi_1-\varphi_2)}{\sin\varphi_2}.
\label{a12-phis}
\end{equation}
If the coefficients satisfy these relations, then the energy \eqref{eq:gl_phase_form} has a critical point at the given $(\varphi_1,\varphi_2)$. 
Next, we check if this critical point is a minimum by calculating the second derivatives of the function \eqref{eq:gl_phase_form}. 
Using Eq.~\eqref{a12-phis}, we obtain that the Hessian matrix is positive-definite if
\begin{equation}
    \sign(a)\frac{\sin\varphi_1}{\sin\varphi_2}<0.
\label{Hessian-positive}
\end{equation}
Taking any point $(\varphi_1,\varphi_2)$ from the stability regions defined by this last inequality and substituting it in Eq.~\eqref{a12-phis}, we find the GL energy for which this pair of phases delivers a minimum (local or global). 

One can easily show that the solutions satisfying Eq.~\eqref{Hessian-positive} exist only if $a\tilde a_1\tilde a_2>0$, i.e., if
\begin{equation*}
    \sign(A_{12}\tilde A_{13}\tilde A_{23})>0.
\end{equation*}
In other words,  we have proved that if the number of negative quadratic inter-component couplings in Eq.~\eqref{F-GL-2} is even, then our system can have TR symmetry-breaking superconducting states, which are at least locally stable. 

Also, we note that the TR symmetry-breaking states with $\varphi_1+\varphi_2=0$ discussed in Sec.~\ref{sec: Bulk spectrum} can only exist at the special values of the coefficients, namely if $\tilde a_1=\tilde a_2$. 
Therefore, such states are unstable against a small variation of the system's parameters, e.g., the temperature.  

\section{Bulk energy spectrum}
\label{app: calculation-E-pm}

Assuming a real $\tilde\eta$, the BdG Hamiltonian \eqref{eq:bdg_matrix_pseudospin_split} in either pseudospin channel can be represented in the form
\begin{equation}
    \hat H=\begin{pmatrix}
        \bnu_1\hbs & \tilde\Delta\hat\sigma_1 \\
        \tilde\Delta^*\hat\sigma_1 & \bnu_2\hbs
    \end{pmatrix},
\label{eq:H-nu-nu}
\end{equation}
where $\bnu_n=(\Re\psi_n,-\Im\psi_n,\xi_n)$. 
It is manifestly particle-hole symmetric and one can find its spectrum either by a direct calculation of a $4\times 4$ determinant \cite{samokhin:2020} or by using the following trick \cite{lo:2022}. 

Let us calculate the second and fourth powers of Eq.~\eqref{eq:H-nu-nu}:
\begin{equation*}
    \hat H^2=\begin{pmatrix}
        \mu_1\hat\sigma_0 & \hat m \\
        \hat m^\dagger & \mu_2\hat\sigma_0,
    \end{pmatrix},
\end{equation*}
where $\mu_n=\nu_n^2+|\tilde\Delta|^2$, $\hat m=\tilde\Delta(\bnu_1\hbs\hat\sigma_1+\bnu_2\hat\sigma_1\hbs)$, and
\begin{equation*}
    \hat H^4=\begin{pmatrix}
        \mu_1^2\hat\sigma_0+\hat m\hat m^\dagger & (\mu_1+\mu_2)\hat m \\
        (\mu_1+\mu_2)\hat m^\dagger &  \mu_2^2\hat\sigma_0+\hat m^\dagger\hat m    
    \end{pmatrix}.
\end{equation*}
One can see that the matrix $\hat M=\hat H^4-(\mu_1+\mu_2)\hat H^2$ does not contain off-diagonal $2\times 2$ blocks. 
Moreover, since 
\begin{equation*}
    \hat m\hat m^\dagger=\hat m^\dagger\hat m=|\tilde\Delta|^2[(\bnu_1-\bnu_2)^2+4\nu_{1,1}\nu_{2,1}]\hat\sigma_0, 
\end{equation*}
we find that $\hat M$ is proportional to the $4\times 4$ unit matrix. 
Therefore, the eigenvalues of $\hat H$ satisfy the following bi-quadratic equation:
\begin{eqnarray*}
    && E^4-(\mu_1+\mu_2)E^2 \\
    && \qquad+\mu_1\mu_2-|\tilde\Delta|^2[(\bnu_1-\bnu_2)^2+4\nu_{1,1}\nu_{2,1}]=0.
\end{eqnarray*}
Solving it, we obtain the Bogoliubov energy branches given by Eq.~\eqref{eq:bdg_eigenvalues}.

\section{BdG formalism in the strip geometry}
\label{app: numerical-BdG}

\begin{figure}[!t]
    \centering
    \includegraphics[width=\linewidth]{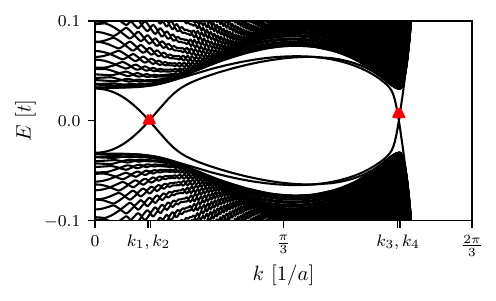}
    \caption{%
    Energy spectrum of $\hat{H}_{\uparrow}(k)$, see Eq.~\eqref{eq:strip_bdg_hamiltonian_spin_split}, in the generic $d_{xy}$-wave case, for the strip geometry at the interband pairing $\tilde{\eta} = 0.4$, see bottom-right plot of Fig.~\ref{fig:periodic_vs_strip_spectrum}. 
    Red triangles: positions $k_1 < k_2 < k_3 < k_4$ for which the corresponding eigenstate profiles are shown in Fig.~\ref{fig:eigenstate_profiles_2}.
    }
    \label{fig:eigenstate_profiles_1}
\end{figure}

\begin{figure*}[!t]
    \centering
    \includegraphics[width=0.85\linewidth]{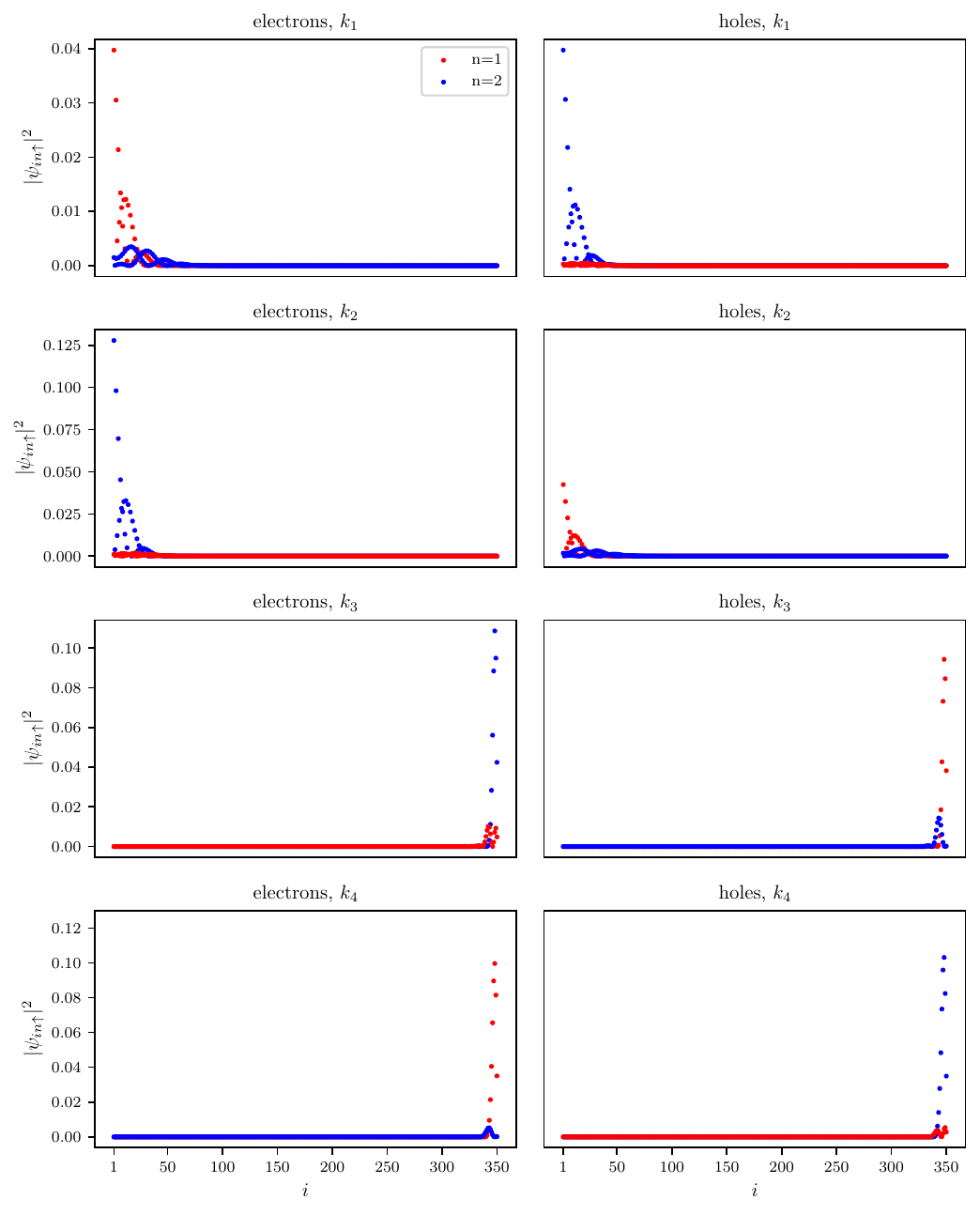}
    \caption{%
    Eigenstate profiles of the states marked by red dots in Fig.~\ref{fig:eigenstate_profiles_1} as a function of $y$ position.
    Left column: electron components of the eigenstates.
    Right column: hole components of the eigenstates.
    }
    \label{fig:eigenstate_profiles_2}
\end{figure*}

The $2N_y\times 2N_y$ matrices $\hat{\xi}_{\uparrow(\downarrow)}$ and $\hat{\Delta}_{\uparrow(\downarrow)}$ appearing in the BdG Hamiltonian \eqref{eq:bdg_hamiltonian_2} are of the form
\begin{equation*}
    \hat{\xi}_{\uparrow(\downarrow)} = \begin{pmatrix}
        \hat{\xi}_{+0\uparrow(\downarrow)} & \hat{\xi}_{+1\uparrow(\downarrow)} & \\
        \hat{\xi}_{-1\uparrow(\downarrow)} & \ddots & \ddots \\
         & \ddots & \ddots
    \end{pmatrix},
\end{equation*}
which is block-tridiagonal, and
\begin{equation*}
    \hat{\Delta}_{\uparrow(\downarrow)} = \begin{pmatrix}
        \hat{\Delta}_{+0\uparrow(\downarrow)} & \hat{\Delta}_{+1\uparrow(\downarrow)} & \hat{\Delta}_{+2\uparrow(\downarrow)} & \\
        \hat{\Delta}_{-1\uparrow(\downarrow)} & \ddots & \ddots & \ddots \\
        \hat{\Delta}_{-2\uparrow(\downarrow)} & \ddots & \ddots &\ddots \\
         & \ddots & \ddots & \ddots
    \end{pmatrix},
\end{equation*}
which is block-quinquediagonal.
The labels $+0, \pm 1, \pm 2$ refer to the respective (off)diagonals of the matrices $\hat{\xi}_{\uparrow(\downarrow)}$ and $\hat{\Delta}_{\uparrow(\downarrow)}$.
Furthermore, regardless of the superconducting pairing we have
\begin{equation*}
\begin{split}
    \hat{\xi}_{+0\uparrow(\downarrow)}
    &= \left[-\frac{\mu_1 + \mu_2}{2} - (t_1 + t_2)\cos(k)\right]\hat\tau_0 \\
    &\peq +\left[-\frac{\mu_1 - \mu_2}{2} - (t_1 - t_2)\cos(k)\right]\hat\tau_3, \\
    \hat{\xi}_{+1\uparrow(\downarrow)}
    &= \left[-\frac{t_1 + t_2}{2} - (t_1' + t_2')\cos(k)\right]\hat\tau_0 \\
    &\peq +\left[-\frac{t_1 - t_2}{2} - (t_1' - t_2')\cos(k)\right]\hat\tau_3,
\end{split}
\end{equation*}
and $\hat{\xi}_{-1\uparrow(\downarrow)}^{\ndagger} = \hat{\xi}_{+1\uparrow(\downarrow)}^{\dagger}$, where $\hat{\tau}_i$ matrices refer to the band space.


Fermionic antisymmetry requires
\begin{equation*}
    \hat{\Delta}_{\uparrow(\downarrow)}(\bk) = \hat{\Delta}_{\downarrow(\uparrow)}^{\top}(-\bk),
\end{equation*}
so that
$\hat{\Delta}_{-1(-2)\uparrow(\downarrow)}(k) = \hat{\Delta}_{+1(+2)\downarrow(\uparrow)}^{\top}(-k)$.
Therefore, in the $s$-wave case, we have
\begin{equation*}
\begin{split}
    \hat{\Delta}_{+0\uparrow(\downarrow)}(k)
    &= \frac{\eta_1 + \eta_2}{2}\hat\tau_0 + \tilde{\eta}\hat\tau_1 + \frac{\eta_1 - \eta_2}{2}\hat\tau_3, \\
    \hat{\Delta}_{+1\uparrow(\downarrow)}(k)
    &= \pm\frac{i}{2}\rho\tilde{\eta}\sin(k)\cos(k)\hat\tau_2, \\
    \hat{\Delta}_{+2\uparrow(\downarrow)}(k),
    &= \mp\frac{i}{4}\rho\tilde{\eta}\sin(k)\hat\tau_2,
\end{split}
\end{equation*}
while in the $d_{xy}$-wave case, we have
\begin{equation*}
\begin{split}
    \hat{\Delta}_{+0\uparrow(\downarrow)}(k)
    &= \mp\rho\tilde{\eta}\cos(k)\hat\tau_2, \\
    \hat{\Delta}_{+1\uparrow(\downarrow)}(k)
    &= \pm\frac{1}{2}\rho\tilde{\eta}\hat\tau_2 - \frac{i}{2}\sin(k) \\
    &\peq\times\left[\frac{\eta_1 + \eta_2}{2}\hat\tau_0 + \tilde{\eta}\hat\tau_1 + \frac{\eta_1 - \eta_2}{2}\hat\tau_3\right], \\
    \hat{\Delta}_{+2\uparrow(\downarrow)}(k)
    &= 0.
\end{split}
\end{equation*}
If periodic boundary conditions are also considered along the $y$ direction (i.e., the strip is closed to a torus), then additional off-diagonal terms appear in the corners of $\hat{\xi}_{\uparrow(\downarrow)}$ and $\hat{\Delta}_{\uparrow(\downarrow)}$.

Fig.~\ref{fig:eigenstate_profiles_1} illustrates that the gap-crossing energy branches in the strong interband-pairing regime belong to states which are localized near the edges of the strip.
The respective eigenstates $\bm{\psi}$ of the Hamiltonian $\hat{H}_{\uparrow}$ are of size $4N_y$ and split into electron and hole contributions as well as band contributions $n = 1$ and $n = 2$, as shown in Fig.~\ref{fig:eigenstate_profiles_2}.

\section{Topological charges of the nodes}
\label{app: nodal charges}

Substituting Eqs.~\eqref{eq:simple-model-bands} and \eqref{eq:simple-model-d-wave-functions} in Eq.~\eqref{eq:det-chiral-block}, we obtain:
\begin{eqnarray}
    \det\hat\upsilon =\left(\frac{\Eb}{2}\right)^2-\tilde\eta^2\left[\rho^2+(1-\rho^2)\sin^2(2\theta)\right]\nonumber\\
    -\xi^2+\eta_1\eta_2\sin^2(2\theta)\nonumber\\
    +i\left[\left(\xi+\frac{\Eb}{2}\right)\eta_1+\left(\xi-\frac{\Eb}{2}\right)\eta_2)\right]\sin(2\theta)
\label{eq:simple-model-det-ups},    
\end{eqnarray}
with real $\eta_1$ and $\eta_2$. Equating the real and imaginary parts of this last expression to zero, we recover the results of Sec.~\ref{sec: d-wave-Phi-nonzero}, if $\eta_1\eta_2<0$, and Sec.~\ref{sec: d-wave-Phi-zero}, if $\eta_1\eta_2>0$.

Let us consider, for example, the high-symmetry nodes along the positive $k_x$ axis, i.e., at $\theta=0$. 
They are located between the two Fermi surfaces, at $\xi=\pm\xi_0$, where 
\begin{equation*}
    \xi_0=\sqrt{\left(\frac{\Eb}{2}\right)^2-\tilde\eta^2\rho^2}.
\end{equation*}
We expand Eq.~\eqref{eq:simple-model-det-ups} near the nodes by setting $\xi=\pm\xi_0+\xi_0 x$, $\theta=y$ ($|x|,|y|\ll 1$), and obtain $\det\hat\upsilon=\mp\xi_0^2(x+iw_\pm y)$,
where
\begin{equation}
   w_\pm=-\frac{1}{\xi_0^2}\left[\left(\xi_0\pm\frac{\Eb}{2}\right)\eta_1+\left(\xi_0\mp\frac{\Eb}{2}\right)\eta_2\right].
\label{eq:w-pm-hs-nodes}
\end{equation}
Therefore, the topological charges of the nodes, see Eq.~\eqref{eq:top-charge-winding}, are given by
\begin{equation*}
    q_\pm=\,\sign(w_\pm).
\end{equation*}
In particular, in the absence of the interband pairing, we have $q_+=-\sign(\eta_1)$ and $q_-=-\sign(\eta_2)$.

It follows from Eq.~\eqref{eq:w-pm-hs-nodes} that
\begin{equation}
    q_+q_-=\,\sign\left[\eta_1\eta_2-\frac{\tilde\eta^2\rho^2}{\Eb^2}(\eta_1+\eta_2)^2\right].
\label{eq:q-plus-q-minus}
\end{equation}
If $\eta_1$ and $\eta_2$ have opposite signs, then $q_+q_-<0$, independently of the value of $\tilde\eta$. 
The high-symmetry nodes on the same axis have opposite charges and annihilate each other at the critical strength of the interband pairing, given by Eq.~\eqref{eq:eta-c-generic-d-wave}.

In contrast, if $\eta_1$ and $\eta_2$ have the same sign, then Eq.~\eqref{eq:q-plus-q-minus} changes sign at
\begin{equation*}
    \tilde\eta_c=\frac{\Eb}{|\rho|}\frac{\sqrt{|\eta_1\eta_2|}}{|\eta_1|+|\eta_2|}.
\end{equation*}
At this value of the interband pairing, one of the high-symmetry nodes splits into two stray nodes of the same charge and one high-symmetry node of the opposite charge, see Sec.~\ref{sec: d-wave-Phi-zero}. 
All nodes eventually annihilate each other at a sufficiently strong interband pairing. 

\end{document}